\documentclass[11pt,a4paper]{article}
\pdfoutput=1

\usepackage{jheppub}
\usepackage{latexsym}
\usepackage{revsymb}
\usepackage{multirow}
\usepackage{color}
\usepackage[usenames,dvipsnames,svgnames,table]{xcolor}

\usepackage{graphicx}
\usepackage{epsfig}  
\usepackage{epsf}    
\usepackage{dcolumn}
\usepackage{bm}
\usepackage{dcolumn}
\usepackage{textcomp}
\usepackage{float}
\usepackage{subfig}
\usepackage{hypcap}
\usepackage[]{hyperref}

\hypersetup{
  bookmarks=true,         
  unicode=false,          
  pdftoolbar=true,        
 pdfmenubar=true,        
 pdffitwindow=true,     
 pdfstartview={FitH},    
 pdfsubject={Neutrino Oscillations Phenomenology},   
 pdfnewwindow=true,      
 pdfcreator={RevTeX},
 colorlinks=true,       
 linkcolor=red,          
 citecolor=blue,        
filecolor=black,      
 urlcolor=blue,           
  }

\def\anue{{\bar\nu_e}}

\def\anumu{{\bar\nu_{\mu}}}

\def\anutau{{\bar\nu_{\tau}}}

\newcommand{\beq}{\begin{equation}}
\newcommand{\eeq}{\end{equation}}
\newcommand{\beqa}{\begin{eqnarray}}
\newcommand{\eeqa}{\end{eqnarray}}

\newcommand{\ty}{{\theta_{13}}}
\newcommand{\tz}{{\theta_{23}}}
\newcommand{\sch}{\sin^2 \theta_{13}}
\newcommand{\stch}{\sin^2 2\theta_{13}}
\newcommand{\sa}{\sin^2 \theta_{23}}

\newcommand{\eps}{\varepsilon}

\newcommand{\ie}{{\it i.e.}}

\preprint{IFIC/14-41, IP/BBSR/2014-10}

\title{Probing Non-Standard Interactions at Daya Bay}

\author[a]{Sanjib Kumar Agarwalla,}
\author[a]{Partha Bagchi,}
\author[b,c]{David V. Forero,} 
\author[b]{Mariam T\'{o}rtola$\,$}

\affiliation[a]{Institute of Physics, Sachivalaya Marg, Sainik School Post, Bhubaneswar 751005, India}
\affiliation[b]{AHEP Group, Institut de F\'{i}sica Corpuscular --
  C.S.I.C./Universitat de Val\`{e}ncia, Parc Cientific \\ de Paterna.
 C/ Catedratico Jos\'e Beltr\'an, 2 E-46980 Paterna (Val\`{e}ncia) - Spain}
 \affiliation[c]{Center for Neutrino Physics, Virginia Tech, Blacksburg, VA 24061, USA}
 
\emailAdd{sanjib@iopb.res.in}
\emailAdd{partha@iopb.res.in}
\emailAdd{dvanegas@ific.uv.es}
\emailAdd{mariam@ific.uv.es}

\abstract 
 {In this article we consider the presence of neutrino non-standard
  interactions (NSI) in the production and detection processes of
  reactor antineutrinos at the Daya Bay experiment.
  We report for the first time, the new constraints on the flavor
  non-universal and flavor universal charged-current NSI parameters,
  estimated using the currently released 621 days of Daya Bay data.
  New limits are placed assuming that the new physics effects are just
  inverse of each other in the production and detection processes.  
  With this special choice of the NSI parameters, we observe a shift 
  in the oscillation amplitude without distorting the $L/E$ pattern of 
  the oscillation probability. This shift in the depth of the oscillation dip 
  can be caused by the NSI parameters as well as by $\theta_{13}$, making it
  quite difficult to disentangle the NSI effects from the standard
  oscillations. We explore the correlations between the NSI parameters
  and $\theta_{13}$ that may lead to significant deviations in the
  reported value of the reactor mixing angle with the help of
  iso-probability surface plots.  Finally, we present the limits on
  electron, muon/tau, and flavor universal (FU) NSI couplings with and
  without considering the uncertainty in the normalization of the
  total event rates. Assuming a perfect knowledge of the event rates
  normalization, we find strong upper bounds $\sim$ 0.1\% for the
  electron and FU cases improving the present limits by one order of
  magnitude. However, for a conservative error of 5\% in the total
  normalization, these constraints are relaxed by almost one order of
  magnitude.}

\arxivnumber{1412.1064}

\begin{document}
\maketitle
\flushbottom

\section{Introduction}
\label{introduction}

The recent discovery of the smallest neutrino mixing angle
$\theta_{13}$ by the modern reactor antineutrino experiments Daya
Bay~\cite{An:2012bu,An:2013zwz,talk-Zhang-Nu2014} and
RENO~\cite{Ahn:2012nd} has firmly established the three-flavor
neutrino
paradigm~\cite{Forero:2014bxa,Capozzi:2013csa,Gonzalez-Garcia:2014bfa}
and signifies an important development towards our understanding of
the structure of the neutrino mass matrix, whose precise
reconstruction would shed light on the underlying new physics that
gives rise to neutrino mass and
mixing~\cite{Mohapatra:2006gs,Morisi:2012fg,King:2014nza}.  Another
reactor electron antineutrino disappearance experiment: Double
Chooz~\cite{Abe:2011fz,Abe:2012tg}, and the two accelerator electron
(anti-)neutrino appearance experiments: MINOS~\cite{Adamson:2013ue}
(completed) and T2K~\cite{Abe:2013hdq,Abe:2013xua} (presently running)
have also confirmed the non-zero and moderately large value of
$\theta_{13}$ in the standard three-flavor oscillation scenario.
It is quite remarkable to see that with 621 days of data taking and
using the merit of identical multi-detector setup, the Daya Bay
experiment reveals a non-zero value of $\theta_{13}$ at more than
16$\sigma$ and suggests a best-fit value of $\stch = 0.084 \pm
0.005$~\cite{talk-Zhang-Nu2014}.  This data provides a relative
1$\sigma$ precision of 6\% on $\stch$ which is already better than the
precision achieved on $\sa$.
Undoubtedly, this high precision measurement of the 1-3 mixing angle
has speeded up the search for the neutrino mass ordering and the
possible presence of a CP-violating phase in current and future
neutrino oscillation
experiments~\cite{Nunokawa:2007qh,Pascoli:2013wca,Agarwalla:2013hma,Agarwalla:2014fva,Minakata:2014tza}.

To explain the presence of small neutrino masses and relatively large
neutrino mixings as indicated by neutrino oscillation data, various
neutrino mass models have been proposed.
These neutrino mass models come in various categories such as the
cases where neutrinos acquire mass via the popular {\it seesaw}
mechanism~\cite{Minkowski:1977sc,Yanagida:1979as,Mohapatra:1979ia,GellMann:1980vs,Schechter:1980gr,Lazarides:1980nt,Mohapatra:1986bd,Akhmedov:1995ip,Akhmedov:1995vm,
Dev:2009aw,Boucenna:2014zba}.
We find also models where neutrinos get mass radiatively due to the
presence of extra Higgs bosons~\cite{Cheng:1980qt,Zee:1980ai,Babu:1988ki} or low
energy supersymmetric hybrid models with spontaneous or bilinear
breaking of R-parity~\cite{Diaz:1997xc,Hirsch:2000ef}. The structure
of the standard electroweak neutral and charged currents gets affected
by the presence of these mechanisms responsible for the neutrino mass
generation~\cite{Schechter:1980gr}.
In most of the cases, in the low energy regime, these effects are
known as non-standard interactions (NSI).  Various extensions of the
Standard Model (SM), such as left-right symmetric models and
supersymmetric models with R-parity violation, predict NSI of
neutrinos with other fermions
\cite{Roulet:1991sm,Guzzo:1991hi,Barger:1991ae,Bergmann:1999pk,Berezhiani:2001rs,Antusch:2008tz,Gavela:2008ra,Malinsky:2008qn,Ohlsson:2009vk}.
The NSI in these models are usually generated via the exchange of new
massive particles at low energies.

Neutrino NSI may be of charged-current (CC) or neutral-current (NC)
type, and they can be classified in two main categories:
flavor-changing NSI, when the flavor of the leptonic current involved
in the process is changed, or flavor-conserving non-universal NSI,
when the lepton flavor is not changed in the process but the strength
of the interaction depends on it, violating the weak universality.  In
the low energy regime, these new interactions may be parameterized in
the form of effective four-fermion Lagrangians:
%
\begin{eqnarray}
  \mathcal{L}_{\rm CC-NSI} \, & = & \,
  \frac{G_{F}}{\sqrt{2}} \, \sum_{f,f^\prime}{\eps}^{s\, d,ff^\prime}_{\alpha\beta} \,
     \left[\bar{\nu}_\beta \gamma^{\rho} (1 - \gamma^{5})  \ell_\alpha \right] \,\,
     \left[\bar{f}^\prime \gamma_{\rho} (1 \pm \gamma^{5}) f \right]\, , 
  \label{eq:CC-NSI-Lagrangian}\\
%
  \mathcal{L}_{\rm NC-NSI} \, & = &  \,
  \frac{G_{F}}{\sqrt{2}} \, \sum_{f}{\eps}^{m,ff}_{\alpha\beta} \,
      \left[\bar{\nu}_\beta \gamma^{\rho} (1 - \gamma^{5}) \nu_\alpha \right] \,\,
      \left[\bar{f}\gamma_{\rho} (1 \pm \gamma^{5}) f \right] \, .
  \label{eq:NC-NSI-Lagrangian}
\end{eqnarray}
%
where $G_F$ is the Fermi constant, $\alpha$ and $\beta$ are neutrino
or lepton flavor indices, $f$ and $f'$ label light SM fermions, and
the dimensionless coefficients $\varepsilon$ parametrize the strength
of the interaction.
Here we denote the CC-NSI couplings as
$\eps^{s\,d,ff^\prime}_{\alpha\beta}$ since they affect in general the
source (s) and detector (d) interactions at neutrino experiments,
while $\eps^{m,ff}_{\alpha\beta}$ refers to the NC-NSI couplings
generally affecting the neutrino propagation in matter (m).
Note that in Eqs.~(\ref{eq:CC-NSI-Lagrangian}) and
(\ref{eq:NC-NSI-Lagrangian}) we have assumed for the new interactions
the same Lorentz structure as for the SM weak interactions,
(V$\pm$A). Even though more general expressions are possible within a
generic structure of operators, as shown in Ref.~\cite{Kopp:2007ne},
these are the dominant contributions for reactor experiments, where we
will focus our attention in this work.

NSI effects may appear at three different stages in a given neutrino
experiment, namely neutrino production, neutrino propagation from the
source to the detector, and neutrino detection. In a short-baseline
reactor experiment, the effects on the neutrino propagation are
negligible since it happens mainly in vacuum. Therefore, the new
generation of short-baseline reactor experiments, such as Daya Bay,
offers an excellent scenario to probe the presence of NSI at the
neutrino production and detection, free of any degeneracy with NSI
propagation effects.
Actually, some work has already been done in the context of NSI at
short-baseline reactor experiments. In particular, forecasts for the
sensitivity of Daya Bay to NSI have been published, for instance, in
Ref.~\cite{Leitner:2011aa}.
More recently, another article presented constraints on neutrino NSI
using the previous Daya Bay data set~\cite{Girardi:2014gna}. The
results derived there are in general agreement with some of the cases
we discuss in Sec.~\ref{NSI-bounds}. Nevertheless, our paper also
studies the phenomenology of some other interesting cases not
considered in Ref.~\cite{Girardi:2014gna}, and provides a detailed
description of the effect of NSI in the neutrino survival
probability. Finally, we also discuss the fragility of the bounds on
the NSI couplings derived using Daya Bay data against the presence of
an uncertainty on the total event rate normalization in the
statistical analysis of reactor data.

Given the production and detection neutrino processes involved in
short-baseline reactor neutrino experiments ($\beta$-decay and inverse
$\beta$-decay), the NSI parameters relevant for these experiments are
$\eps_{e\alpha}^{ud}$, i.e., the CC-NSI couplings between up and down
quarks, positrons and antineutrinos of flavor $\alpha$. In the
literature we can find the following 90\% C.L. bounds on these
parameters~\cite{Biggio:2009nt}:
\begin{equation}
|\eps_{e\alpha}^{ud, V}| < 0.041 \,, \quad |\eps_{e\mu}^{ud, L}| < 0.026  \, , \quad  |\eps_{e\mu}^{ud, R}| < 0.037 \,,
\label{eq:bound-biggio}
\end{equation}
coming from unitarity constraints on the CKM matrix as well as from
the non-observation of neutrino oscillations in the NOMAD experiment.
Here the couplings with different chirality 
are related by\footnote{The NSI parameters probed in
  our analysis get contributions from the (V$\pm$A) operators in
  Eq.~(\ref{eq:CC-NSI-Lagrangian}) and therefore they can be generally
  expressed as $\eps_{\alpha\beta}^{ud, \text{V$\pm$A}}$ or
  equivalently $\eps_{\alpha\beta}^{ud, \text{L$\pm$R}}$. However, for
  simplicity, we have dropped the chirality indices all over the
  paper.}:
$\eps_{\alpha\beta}^{ff', V} = \eps_{\alpha\beta}^{ff', L} +
\eps_{\alpha\beta}^{ff', R}$.
Other constraints on neutrino NSI couplings using solar and reactor
neutrinos have been given
in~\cite{Berezhiani:2001rt,Miranda:2004nb,Barranco:2005ps,Barranco:2007ej,Bolanos:2008km,Escrihuela:2009up}.
On the other hand, NSI have also been studied in the context of
laboratory experiments with accelerators in
Refs.~\cite{Davidson:2003ha,Barranco:2007ej,Escrihuela:2011cf,Forero:2011zz},
while bounds from atmospheric neutrino data have been presented in
Refs.~\cite{Fornengo:2001pm, Gonzalez-Garcia:2013usa}.  Recently a
forecast for the sensitivity to NSI of the future PINGU detector has
been presented in \cite{Choubey:2014iia}.
Future medium-baseline reactor experiments like JUNO can also serve as test 
bed to look for NSI~\cite{Li:2014mlo}.

This paper is organized as follows.  In Sec.~\ref{Reactor-NSI}, we
describe the procedure for implementing the NSI effect in the modern
reactor experiments.  There, we derive the effective antineutrino
survival probability expressions which we use later to analyze the
Daya Bay data. We also give plots to discuss in detail the impact of
the NSI parameters on the effective probability considering a special
case where $\varepsilon^s_{e\gamma} = \varepsilon^{d*}_{\gamma e}$.
We also show the possible correlations between the NSI parameters and
$\theta_{13}$ with the help of iso-probability surface plots.
Sec.~\ref{Analysis-DayaBay} describes the numerical methods adopted to
analyze the reactor data. Apart from this, a brief description of the
Daya Bay experiment and the important features of its present data set
which are relevant for the fit are also given in this
section. Sec.~\ref{NSI-bounds} presents the constraints on the NSI
parameters imposed by the current Daya Bay data assuming a perfect
knowledge of the event rates normalization. Next we derive the bounds
on the NSI parameters taking into account the uncertainty in the
normalization of event rates with a prior of 5\% in
Sec.~\ref{NSI-bounds-freenorm}.
In Sec.~\ref{Sec-stat-Daya-Bay}, we compare the constraints on the NSI
parameters obtained with the current 621 days of Daya Bay data with
the limits derived using the previously released 217 days of Daya Bay
data.
Finally, we summarize and draw our conclusions in
Sec.~\ref{Sec-Summary}.  In Appendix~\ref{appendix1}, we give the
effective antineutrino survival probability expressions for the
physical situations where $\varepsilon^s_{e\gamma} \ne
\varepsilon^{d*}_{\gamma e}$.

\section{Implementing NSI in Modern Reactor Experiments}
\label{Reactor-NSI}

According to the usual procedure followed in the non-standard analyses
of reactor data~\cite{Ohlsson:2008gx,Leitner:2011aa}, we start by
re-defining the neutrino flavour states in the presence of NSI in the
source and detection processes. For the initial (at source) and final
(at detector) neutrino flavor states, we have~\cite{Bilenky:1992wv,
  Grossman:1995wx,GonzalezGarcia:2001mp,Meloni:2009cg}:
\begin{equation}\label{eq:nu-def}
 |\nu_\alpha^s \rangle = \frac{1}{N^s_\alpha} \left(|\nu_\alpha \rangle+\sum_\gamma \varepsilon^s_{\alpha \gamma}|\nu_\gamma \rangle \right) , \quad
\langle \nu_\beta^d | = \frac{1}{N^d_\beta} \left(\langle\nu_\beta |+\sum_\eta \varepsilon^d_{\eta \beta}\langle\nu_\eta |\right) ,
\end{equation}
while the redefinition of the antineutrino flavor states is given by:
\begin{equation}\label{eq:nubar-def}
 |\bar{\nu}_\alpha^s \rangle = \frac{1}{N^s_\alpha} \left(|\bar{\nu}_\alpha \rangle+\sum_\gamma \varepsilon^{s *}_{\alpha \gamma}|\bar{\nu}_\gamma \rangle \right), \quad
\langle \bar{\nu}_\beta^d | =  \frac{1}{N^d_\beta} \left( \langle \bar{\nu}_\beta |+\sum_\eta \varepsilon^{d *}_{\eta \beta}\langle\bar{\nu}_\eta | \right).
 \end{equation}
 The normalization factors required to obtain an orthonormal basis can
 be expressed as:
\begin{equation}\label{eq:norm-fact}
 N^s_\alpha = \sqrt{\left[(1+\varepsilon^s)(1+\varepsilon^{s \dagger}) \right]_{\alpha \alpha}} \, , \quad 
 N^d_\beta = \sqrt{\left[(1+\varepsilon^{d \dagger})(1+\varepsilon^d) \right]_{\beta \beta}},
 \end{equation}
 and the neutrino mixing between flavor and mass eigenstates is given
 by the usual expressions:
\begin{equation}\label{eq:nu-mixing}
|\nu_\alpha \rangle=\sum_{k} U^*_{\alpha k}|\nu_k \rangle , \quad
|\bar{\nu}_\alpha \rangle=\sum_{k} U_{\alpha k}|\bar{\nu}_k \rangle\,.
\end{equation}

The correct normalization of the neutrino states in presence of NSI is
a very important point, required to obtain a total neutrino transition
probability normalized to 1.  However, one has to consider that when
dealing with a non-orthonormal neutrino basis, the normalization of
neutrino states will affect not only the neutrino survival probability
but also the calculation of the produced neutrino fluxes and detection
cross sections. In this case, as shown in Ref.~\cite{Antusch:2006vwa},
all the normalization terms coming from $N_\alpha^s$ and $N_\beta^d$
will cancel while convoluting the neutrino oscillation probabilities,
cross sections, and neutrino fluxes to estimate the number of events
in a given experiment such as Daya Bay. This is due to the fact that
the SM cross sections and neutrino fluxes used in our simulation have
been theoretically derived assuming an orthonormal neutrino basis and
therefore they need to be corrected.
Then, from here we can consider the following effective redefinition of neutrino and antineutrino states:
\begin{equation}\label{eq:nu-def2}
 |\nu_\alpha^s \rangle_{\rm eff} =|\nu_\alpha \rangle+\sum_\gamma \varepsilon^s_{\alpha \gamma}|\nu_\gamma \rangle , \quad
\langle \nu_\beta^d |_{\rm eff}  =\langle\nu_\beta |+\sum_\eta \varepsilon^d_{\eta \beta}\langle\nu_\eta | ,
 \end{equation}
\begin{equation}\label{eq:nubar-def2}
 |\bar{\nu}_\alpha^s \rangle_{\rm eff}  = |\bar{\nu}_\alpha \rangle+\sum_\gamma \varepsilon^{s *}_{\alpha \gamma}|\bar{\nu}_\gamma \rangle , \quad
\langle \bar{\nu}_\beta^d |_{\rm eff}  = \langle \bar{\nu}_\beta |+\sum_\eta \varepsilon^{d *}_{\eta \beta}\langle\bar{\nu}_\eta | ,
 \end{equation}
 where we have dropped the normalization factors that will cancel in
 the Monte Carlo simulation of Daya Bay data. From these effective
 neutrino states we will calculate an {\it effective} neutrino
 oscillation probability, that will be used all along our analysis.
 Note that when we will discuss the features of the probability prior
 to the simulation of a particular experiment, we will always refer to
 the {\it effective probability}, that might be greater than one.

\subsection{Effective antineutrino survival probability in reactor experiments}
\label{Effective-NSI-Probability}

The effective antineutrino transition probability from flavor $\alpha$
to $\beta$ after traversing a distance $L$ from source to detector is
defined as:
\begin{equation}
P_{\bar{\nu}_\alpha^s \to \bar{\nu}_\beta^d} =|\langle \bar{\nu}_\beta^d |\exp{(-i\,H\,L)}|\bar{\nu}_\alpha^s \rangle|^2 .
\label{eq:05}
\end{equation}
In terms of the neutrino mass differences and mixing angles, this
transition probability in vacuum may be written as:
\begin{equation}
\begin{split}
P_{\bar{\nu}_\alpha^s \to \bar{\nu}_\beta^d}&=\sum_{j,k} Y^j_{\alpha \beta} Y^{k *}_{\alpha \beta}-4\sum_{j>k} \mathcal{R}\{Y^j_{\alpha \beta} Y^{k *}_{\alpha \beta}\}\,\sin^2 \left(\frac{\Delta m^2_{jk}\,L}{4E}\right)\\
&+2\sum_{j>k} \mathcal{I}\{Y^j_{\alpha \beta} Y^{k *}_{\alpha \beta}\}\,\sin \left(\frac{\Delta m^2_{jk}\,L}{2E}\right),
\end{split}
\label{eq:06}
\end{equation}
where $\Delta m^2_{jk} = m^2_{j} - m^2_{k}$.  In the case of standard
oscillations, $Y^j_{\alpha \beta}$ is defined as:
\begin{equation}
Y^j_{\alpha \beta} \equiv U^*_{\beta j} U_{\alpha j}.
\label{eq:07a}
\end{equation}
In presence of NSI, however, according to the definition of neutrino
states in Eq.~(\ref{eq:nubar-def}), this expression is modified as
follows~\cite{Ohlsson:2008gx}:
\begin{equation}
Y^j_{\alpha \beta} \equiv U^*_{\beta j} U_{\alpha j}+\sum_{\gamma} \varepsilon^{s *}_{\alpha \gamma} U^*_{\beta j} U_{\gamma j}+\sum_{\eta} \varepsilon^{d *}_{\eta \beta} U^*_{\eta j} U_{\alpha j}+\sum_{\gamma, \eta} \varepsilon^{s *}_{\alpha \gamma} \varepsilon^{d *}_{\eta \beta} U^*_{\eta j} U_{\gamma j}.
\label{eq:07}
\end{equation}
To obtain the $\bar\nu_e$ survival probability in a reactor
experiment, where an electron antineutrino is produced at the source
and a positron is detected inside the detector, one has to replace
$\alpha$ and $\beta $ by e in Eq.~(\ref{eq:06}).
For the NSI parameters, we adopt the following parametrization by
splitting the new couplings into its absolute value and its phase:
\begin{equation}
\varepsilon^s_{e\gamma} \equiv |\varepsilon^s_{e\gamma}| \, {\rm e}^{{\rm i} \phi^s_{e\gamma}} \,\,\,\,\,
{\rm and} \,\,\,\,\,
\varepsilon^d_{\eta e} \equiv |\varepsilon^d_{\eta e}| \, {\rm e}^{{\rm i} \phi^d_{\eta e}}.
\label{eq:08}
\end{equation}
Now expanding the various terms of the general transition probability
as given in Eq.~(\ref{eq:06}) and using the parametrization above, we
obtain the effective $\bar\nu_e$ survival probability:
\begin{eqnarray}
\label{eq:pee_tot}
 P_{\bar{\nu}_e^s \to \bar{\nu}_e^d} & =   & P^{\text{SM}}_{\bar{\nu}_e \to
  \bar{\nu}_e}+ P_{\text{non-osc}}^{\text{NSI}} +  P_{\text{osc-atm}}^{\text{NSI}}  + P_{\text{osc-solar}}^{\text{NSI}}  \\
   &+ & \mathcal{O}\left[\eps^3,s_{13}^3, \eps^2 s_{13},  \eps s_{13}^2, \eps s_{13}\left(\frac{\Delta m^2_{21}\,L}{2E}\right), \eps\left(\frac{\Delta m^2_{21}\,L}{2E}\right)^2,s_{13}^2\left(\frac{\Delta m^2_{21}\,L}{2E}\right) \right], \nonumber
\end{eqnarray}
where the Standard Model (SM) contribution is given by
\begin{equation}
  P^{\text{SM}}_{\bar{\nu}_e \to \bar{\nu}_e} 
 = 1-\sin^2 2\theta_{13}\left(c^2_{12}\sin^2 \Delta_{31} + s^2_{12} \sin^2 \Delta_{32} \right)
-  c^4_{13}\sin^2 2\theta_{12} \sin^2\Delta_{21},
\label{eq:pee_sm}
\end{equation}
with $s_{ij} = \sin\theta_{ij}$, $c_{ij} = \cos\theta_{ij}$, and
$\Delta_{ij} = \Delta m^2_{ij} L/4E$.  The various NSI terms of the
effective survival probability in Eq.~(\ref{eq:pee_tot}) can be
written as:
\begin{eqnarray}
P_{\text{non-osc}}^{\text{NSI}}  & =&  2 \left (|\varepsilon^d_{ee}|
  \cos\phi^d_{ee}+|\eps^s_{ee}| \cos\phi^s_{ee} \right) + |\eps^{d}_{ee}|^2+|\eps^{s}_{ee}|^2 +  2 |\eps^d_{ee}|
 |\eps^s_{ee}| \cos(\phi^d_{ee}-\phi^s_{ee}) \label{eq:pee_nonosc} \\ \nonumber 
& + & 2 |\eps^d_{ee}|
 |\eps^{s}_{ee}| \cos(\phi^d_{ee}+\phi^s_{ee}) + 2 |\eps^s_{e\mu}| |\eps^d_{\mu e}| \cos(\phi^s_{e
   \mu}+\phi^d_{\mu e})+2 |\eps^s_{e\tau}| |\eps^d_{\tau e}|
 \cos(\phi^s_{e \tau}+\phi^d_{\tau e}), \\
 P_{\text{osc-atm}}^{\text{NSI}} & = & 2 \left\{ 
 s_{13} s_{23} \left[|\eps^s_{e \mu}| \sin(\delta-\phi^s_{e \mu}) - |\eps^d_{\mu e}| \sin(\delta+\phi^d_{\mu e})\right]
  \right . \nonumber \\
& + & \left . s_{13} c_{23} \left[ |\eps^s_{e\tau}| \sin(\delta-\phi^s_{e \tau}) -|\eps^d_{\tau e}| \sin(\delta+\phi^d_{\tau e})\right]
  \right . \nonumber \\
 &- &\left . s_{23} c_{23} \left[ |\eps^s_{e \mu}| |\eps^d_{\tau e}| \sin(\phi^s_{e
   \mu}+\phi^d_{\tau e}) + |\eps^s_{e \tau}|
  |\eps^d_{\mu e}| \sin(\phi^s_{e \tau}+\phi^d_{\mu e}) \right]
\right . \nonumber  \\
  &-& \left .  c_{23}^2 |\eps^s_{e \tau}| |\eps^d_{\tau e}| 
  \sin(\phi^s_{e \tau}+ \phi^d_{\tau e})
  - s_{23}^2 |\eps^s_{e \mu}| |\eps^d_{\mu
   e}| \sin(\phi^s_{e \mu}+\phi^d_{\mu e})  \right \} \,\sin \left(2\Delta_{31}\right)  \nonumber \\
 & - & 4 \left\{  s_{13} s_{23}\left[
 |\eps^s_{e\mu}|  \cos(\delta-\phi^s_{e\mu})   +|\eps^d_{\mu e}|  \cos(\delta+\phi^d_{\mu e}) \right]    \right .\nonumber \\
  & + & \left .   s_{13} c_{23} \left[
 |\eps^s_{e\tau}|  \cos(\delta-\phi^s_{e \tau})
 +   |\eps^d_{\tau e}|  \cos(\delta+\phi^d_{\tau e}) \right]
\right . \nonumber \\
& + & \left . s_{23} c_{23}\left[  |\eps^s_{e\mu}| |\eps^d_{\tau e}| \cos(\phi^s_{e\mu}+\phi^d_{\tau e}) +
|\eps^s_{e \tau}| |\eps^d_{\mu e}| \cos(\phi^s_{e \tau}+\phi^d_{\mu e})  \right] \right . \nonumber  \\ 
 &+ & \left 	.  
 c_{23}^2 |\eps^s_{e \tau}| |\eps^d_{\tau e}| \cos(\phi^s_{e\tau}+\phi^d_{\tau e}) +
   s_{23}^2 |\eps^s_{e \mu}| |\eps^d_{\mu e}| \cos(\phi^s_{e \mu}+\phi^d_{\mu e})  \right\} \sin^2 \left(\Delta_{31}\right),  
   \label{eq:pee_oscatm} \\
 P_{\text{osc-solar}}^{\text{NSI}}& = & 2 \sin 2\theta_{12} \Delta_{21} \, \left \{
 -c_{23} (|\eps^s_{e \mu}| \sin\phi^s_{e\mu}+|\eps^d_{\mu e}| \sin\phi^d_{\mu e}) \right .\nonumber \\
 & + & \left . s_{23}(|\eps^s_{e \tau}| \sin\phi^s_{e\tau} +|\eps^d_{\tau e}| \sin\phi^d_{\tau e})  \right \}. 
  \label{eq:pee_oscsolar}
\end{eqnarray}
The linear coefficients of the terms of order $|\eps|$ in
Eqs.~(\ref{eq:pee_nonosc}, \ref{eq:pee_oscatm}, \ref{eq:pee_oscsolar})
are the same as given in Ref.~\cite{Kopp:2007ne}, with a good
agreement between the calculated probabilities here and there.
However, as it can be seen in the expressions above, here  
we also include new terms up to second order in  $|\eps|$ in the  effective neutrino probability. 
The relevance of these corrections will be discussed later in the paper.
In Eq.~(\ref{eq:pee_oscatm}), note the presence of a term linear in
the sine of $\Delta m^2_{31} L/2E$ which therefore depends on the
choice of neutrino mass ordering. This term does not appear in the
standard $\bar\nu_e \to \bar\nu_e$ oscillation expression and it can
affect the $L/E$ dependence of the probability in the presence of
neutrino NSI.

\subsection{Special case of NSI: $\varepsilon^s_{e\gamma} = \varepsilon^{d*}_{\gamma e}$}
\label{Special-NSI-Case}

In this work\footnote{In the appendix, we have given the effective
  probability expressions for the physical situations where
  $\varepsilon^s_{e\gamma} \ne \varepsilon^{d*}_{\gamma e}$. In such
  cases, the spectral analysis of the reactor data becomes inevitable
  since the NSI parameters not only cause a shift in $\theta_{13}$
  {\ie} the change of the {\it depth} of the first oscillation maximum
  but also modify the $L/E$ pattern of the oscillation probability
  due to the shift in its energy. A detailed analysis of the Daya Bay data 
  under such scenarios will be performed in \cite{shape-nsi-reactor}.} we assume
that, likewise the mechanisms responsible for production (via
$\beta$-decay) and detection (via inverse $\beta$-decay) of reactor
antineutrinos are just inverse of each other, this is also true for
the associated NSI~\cite{Kopp:2007ne,Leitner:2011aa}.
This assumption allows us to write $\varepsilon^s_{\gamma} =
\varepsilon^{d*}_{\gamma} \equiv |\varepsilon_{\gamma}| {\rm e}^{{\rm
    i} \phi_{\gamma}}$ where we drop the universal e index for
simplicity.  With these assumptions, Eq.~(\ref{eq:pee_tot}) takes the
form (keeping the terms up-to the second order in small quantities):
\begin{eqnarray}
P_{\bar{\nu}^s_e \rightarrow \bar{\nu}^d_e} &\simeq& 
{\underbrace{
1-\sin^2 2\theta_{13}\left(c^2_{12}\sin^2 \Delta_{31} + s^2_{12} \sin^2 \Delta_{32} \right)
-  c^4_{13}\sin^2 2\theta_{12} \sin^2\Delta_{21}}_{\rm{Standard~Model~terms}}} \nonumber \\
&+& 
{\underbrace{4|\varepsilon_e| \text{cos}{\phi_e} + 4 |\varepsilon_e|^2  + 2|\varepsilon_e|^2
\cos{2 \phi_e} + 2 |\varepsilon_{\mu}|^2 + 2 |\varepsilon_{\tau}|^2}_{\rm{non-oscillatory~NSI~terms}}} \nonumber  \\ 
&-&
{\underbrace{4\{ s^2_{23} |\varepsilon_{\mu}|^2 + c^2_{23} |\varepsilon_{\tau}|^2  + 
2s_{23}c_{23}|\varepsilon_{\mu}| |\varepsilon_{\tau}|\text{cos}{(\phi_{\mu} - \phi_{\tau})}\}
\sin^2 \Delta_{31}}_{\rm{oscillatory~NSI~terms}}} \nonumber \\
&-& 
{\underbrace{4\{2s_{13}[s_{23}|\varepsilon_{\mu}|\cos{(\delta-\phi_{\mu}) +
c_{23}|\varepsilon_{\tau}|\cos(\delta-\phi_{\tau})]\}
\sin^2\Delta_{31}}}_{\rm{oscillatory~NSI~terms}}}.
\label{eq:case1}
\end{eqnarray}
For this special case of NSI parameters, there is no linear
sine-dependent term in Eq.~(\ref{eq:case1}) and two striking features
are emerging from the effective probability expression which are
responsible for a change in the oscillation amplitude. First we can see
the presence of some non-oscillatory NSI terms which are independent
of $L$ and $E$ and are given by
\begin{eqnarray}
1+ 4|\varepsilon_e| \cos{\phi_e} + 4 |\varepsilon_e|^2  + 2|\varepsilon_e|^2 \cos{2 \phi_e} + 2 |\varepsilon_{\mu}|^2 + 2 |\varepsilon_{\tau}|^2 \,,
\label{eq:normalization-shift}
\end{eqnarray}
and, second, there is a shift in the {\it effective} 1-3 mixing angle due to oscillatory NSI terms which can be written as  
\begin{eqnarray}
s_{13}^2  & \to & s_{13}^2+s^2_{23} |\varepsilon_{\mu}|^2 + c^2_{23} |\varepsilon_{\tau}|^2  + 2s_{23}c_{23}|\varepsilon_{\mu}| 
|\varepsilon_{\tau}|\cos(\phi_\mu - \phi_\tau) \nonumber \\
& + & 2s_{13}\left[s_{23}|\varepsilon_{\mu}|\cos(\delta-\phi_\mu) + c_{23}|\varepsilon_\tau| \cos(\delta-\phi_\tau)\right].
\label{eq:th13-shift}          
\end{eqnarray}
These two features, which are brought about by the NSI parameters, are
responsible for a shift in the oscillation amplitude without distorting the 
$L/E$ pattern of the oscillation probability as can be clearly
seen from Fig.~\ref{fig:prob-band-1} and Fig.~\ref{fig:prob-band-2}
that we will discuss in the next section.  Eq.~(\ref{eq:th13-shift})
suggests that it will be quite challenging to discriminate the effect
of {\it true} $\theta_{13}$ and NSI parameters in the modern reactor
experiments. It is also interesting to note that there are some CP
conserving terms in Eq.~(\ref{eq:th13-shift}) which come into the
picture due to the presence of NSI parameters. One of the most
important consequences of the new non-oscillatory NSI terms (see
Eq.~(\ref{eq:normalization-shift})) is that they can cause a flavor
transition at the source ($L$ = 0) even before neutrinos start to
oscillate. In the literature, this feature is known as
``zero-distance'' effect~\cite{Kopp:2007ne, Langacker:1988up}.
In modern reactor experiments, this effect can be probed using the
near detectors which are placed quite close to the source.

For definiteness, in this work we have restricted our analysis to the
following choices of the NSI parameters:

\begin{itemize}

\item Lepton number conserving {\it non-universal} NSI parameters
  which depend on the flavor characterizing the violation of weak
  universality.  Under this category, we study the following two
  cases:

\begin{enumerate}

\item Considering only the NSI parameters $|\varepsilon_e|$ and
  $\phi_e$ which are associated with $\anue$.  In the presence of
  these flavor {\it conserving} NSI parameters, Eq.~(\ref{eq:case1})
  takes the form:
\begin{equation}
P_{\bar{\nu}^s_e \rightarrow \bar{\nu}^d_e}^{\text{NSI-e}} \simeq P^{\text{SM}}_{\bar{\nu}_e \to \bar{\nu}_e} +
4|\varepsilon_e| \text{cos}{\phi_e} + 4 |\varepsilon_e|^2  + 2|\varepsilon_e|^2 \cos{2 \phi_e}.
\label{eq:only-nuebar}
\end{equation}
At first order in $|\varepsilon_e|$ and neglecting the effect of the
solar mass splitting, the new non-oscillatory NSI terms appearing at
the survival probability produce a total shift in the effective
$\theta_{13}$ mixing angle given by:
\begin{equation}
\tilde{s}^2_{13} \approx s^2_{13}-\frac{|\varepsilon_e| 
\text{cos}{\phi_e}}{\sin^2{\Delta_{31}}}\,.
\label{eq:shift-e}
\end{equation}
This expression will be very useful to discuss the behavior of the
effective probability as well as the correlations between
$\theta_{13}$ and the NSI parameters in the next subsections.

\item Considering only the NSI parameters $|\varepsilon_\mu|$ and
  $\phi_\mu$ which are associated with $\anumu$.  In the presence of
  these flavor {\it violating} NSI parameters, Eq.~(\ref{eq:case1})
  takes the form:
\begin{equation}
P_{\bar{\nu}^s_e \rightarrow \bar{\nu}^d_e}^{\text{NSI-}\mu} \simeq 
P^{\text{SM}}_{\bar{\nu}_e \to \bar{\nu}_e} +
2 |\varepsilon_{\mu}|^2 -
4 \{s^2_{23} |\varepsilon_{\mu}|^2 + 2s_{13}s_{23}|\varepsilon_{\mu}|\cos(\delta-\phi_{\mu}) \}
\sin^2\Delta_{31}.
\label{eq:only-numubar}
\end{equation}
Note that, for the NSI parameters $|\varepsilon_\tau|$ and $\phi_\tau$
which are associated with $\anutau$, the effective survival
probability will be exactly the same as Eq.~(\ref{eq:only-numubar})
with the replacements $|\varepsilon_{\mu}| \to |\varepsilon_{\tau}|$
and $\phi_{\mu} \to \phi_{\tau}$ provided that the 2-3 mixing angle is
maximal ,\ie, $\sa = 0.5$.
As before, the new non-standard oscillatory terms in the neutrino
transition probability may be interpreted as a global redefinition of
the effective $\theta_{13}$ mixing angle:
\begin{equation}
\tilde{s}^2_{13} \approx 
s^2_{13}+2s_{13}s_{23}|\varepsilon_{\mu}|\cos{(\delta-
\phi_{ \mu})}.
\label{eq:shift-mu}
\end{equation} 

\end{enumerate}

\item Lepton number conserving {\it universal} NSI parameters which do
  not depend on flavor.  In this case, we have $|\eps_e| = |\eps_\mu|
  = |\eps_\tau| = |\eps|$ and $\phi_e = \phi_\mu = \phi_\tau = \phi$
  and the probability in Eq.~(\ref{eq:case1}) takes the form:
\begin{eqnarray}
P_{\bar{\nu}^s_e \rightarrow \bar{\nu}^d_e}^{\text{NSI-$\alpha$}} &\simeq& 
P^{\text{SM}}_{\bar{\nu}_e \to \bar{\nu}_e} +
4|\eps| \text{cos}{\phi} + 2 |\eps|^2  \left(4 + \cos{2 \phi}\right) \nonumber \\ 
&-& 4  \{|\eps|^2 + 2s_{23}c_{23}|\eps|^2 + 2s_{13}|\eps|\cos(\delta-\phi)(s_{23} + c_{23})\}\sin^2\Delta_{31}.
\label{eq:flavorless}
\end{eqnarray}
In this case, the effective mixing angle in the presence of
oscillatory and non-oscillatory NSI terms will be given by:
 \begin{equation}
\tilde{s}^2_{13} 
\approx s^2_{13}-|\varepsilon|\left[\frac{\text{cos}{\phi}}{\sin^2{\Delta_{31}}}
-2s_{13}(s_{23}+c_{23} )\cos{(\delta-\phi)}\right].
\label{eq:shift-univ}
\end{equation} 

\end{itemize}

As stated above, the expressions given in this subsection to  illustrate the shift in the effective 
reactor angle in the presence of NSI, Eqs.~(\ref{eq:shift-e}), (\ref{eq:shift-mu}) and 
(\ref{eq:shift-univ}), contain only first order corrections in the NSI couplings. Note, however, 
that terms of second order in $|\varepsilon|$ have been included in all the numerical results 
shown along the paper.
Here we will briefly discuss the relevance of second order corrections in our analysis.
Clearly, these corrections to the effective neutrino probability are only significant
 in the cases where first order corrections are very small or totally cancelled. This happens for 
 $\phi_e = \pm90^\circ$ in the case of electron-NSI couplings, for ($\delta -
\phi_{\mu,\tau}) = \pm 90^\circ$ in the case of muon/tau-NSI couplings
and for ($\delta=0$, $\phi=\pm90^\circ$) in the flavour-universal case.
 From Eqs.~(\ref{eq:only-nuebar}), (\ref{eq:only-numubar}) and (\ref{eq:flavorless}),
 it is straightforward to evaluate the size of the second order terms for the three cases under study.
Taking into account the baselines and neutrino energies probed at the Daya Bay experiment, we find that 
 second order corrections (for vahishing first order corrections)  are approximately given by: 
$0.5|\varepsilon_e|^2$ (electron-NSI case), 
$0.05|\varepsilon_{\mu,\tau}|^2$ (muon/tau-NSI case) and $0.4|\varepsilon|^2$ (flavour-universal-NSI case).
The small size of the corrections in the muon/tau-NSI, one order of magnitude smaller than for the two other cases, comes from the smallness of 
 the coefficient responsible for second order corrections in the
expression of the effective reactor angle, $\left(\frac{1}{2 \sin^2 \Delta_{31}}-s^2_{23}\right)$, very close
to zero for the energies and baselines studied in Daya Bay. 
 This result can be observed in the correlation plots in Figs.~\ref{fig:isoprob-1}  and \ref{fig:isoprob-2} (see dashed
blue line) in Sec.~\ref{iso-probability}. There, one sees that second order corrections are small but visible for electron-NSI 
and flavour-universal case, while they are almost negligible for the muon/tau-NSI case. 
The impact of the second order corrections over the results presented in this work is discussed in Sec.~\ref{Sec-Summary}.

\subsection{Impact of the NSI parameters on the effective probability} 
\label{probability-vs-energy}

\begin{table}[!tb]
\centering
\begin{tabular}{|l|l|l|l|l|l|l|}
\hline
Parameter & $\sin^2\theta_{12}$ & $\sin^2\theta_{23}$ & $\sin^2\theta_{13}$ &  $\Delta m_{21}^2 (\rm eV^2)$ & $\Delta m_{31}^2 (\rm eV^2)$ & $\delta$ \\
\hline
Value & 0.32 & 0.5 & 0.023 & 7.6 $\times$ $10^{-5}$ & 2.55 $\times \: 10^{-3}$ & 0 -- 2$\pi$ \\
\hline
\end{tabular}
\caption{Benchmark values of the neutrino oscillation parameters used in this work, taken from Refs.~\cite{Tortola:2012te,Forero:2014bxa}.}
\label{tab_osc_param_input}
\end{table}

\begin{figure}[!tb]
\centering
\includegraphics[width=0.49\textwidth]{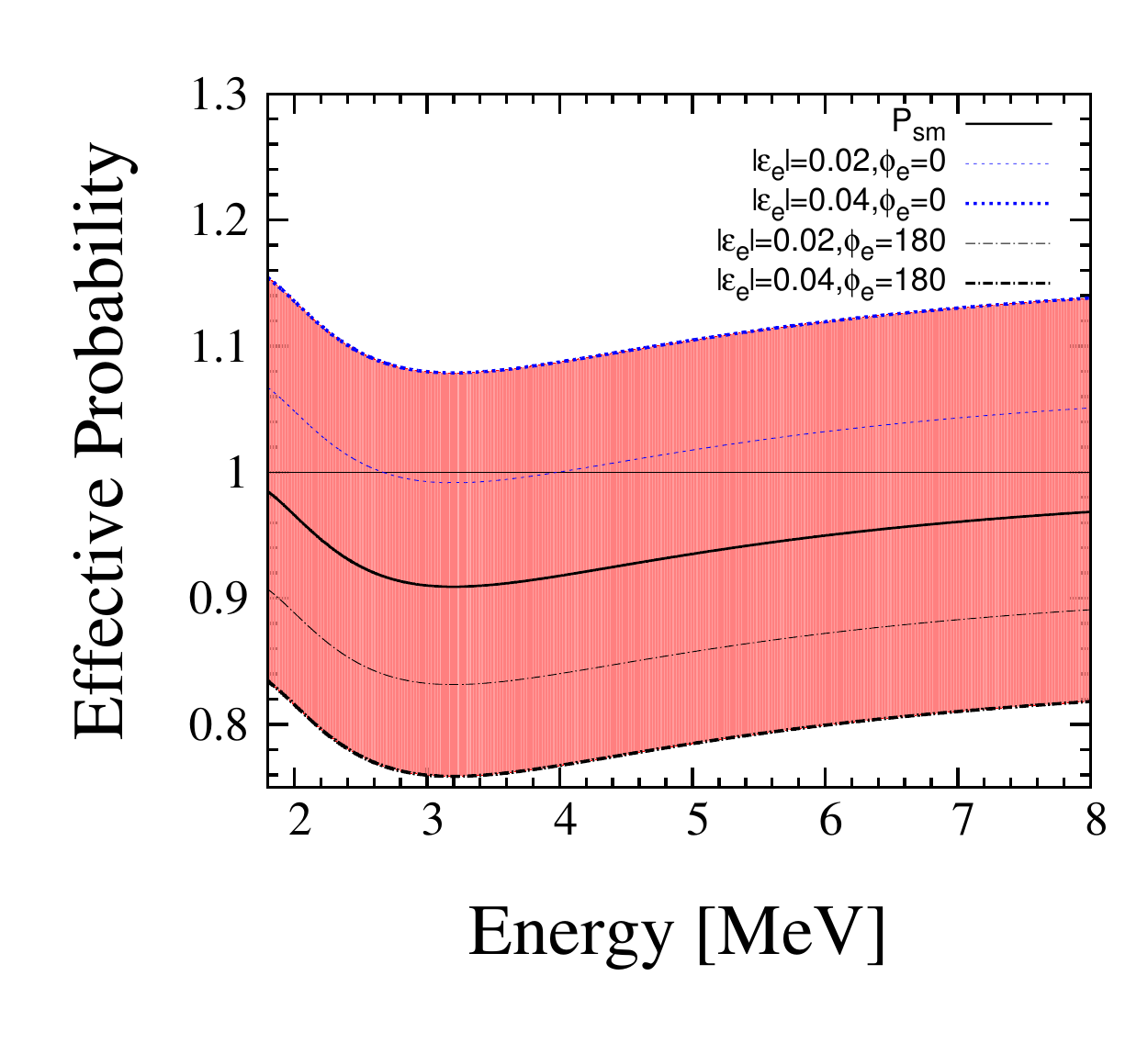}
\includegraphics[width=0.49\textwidth]{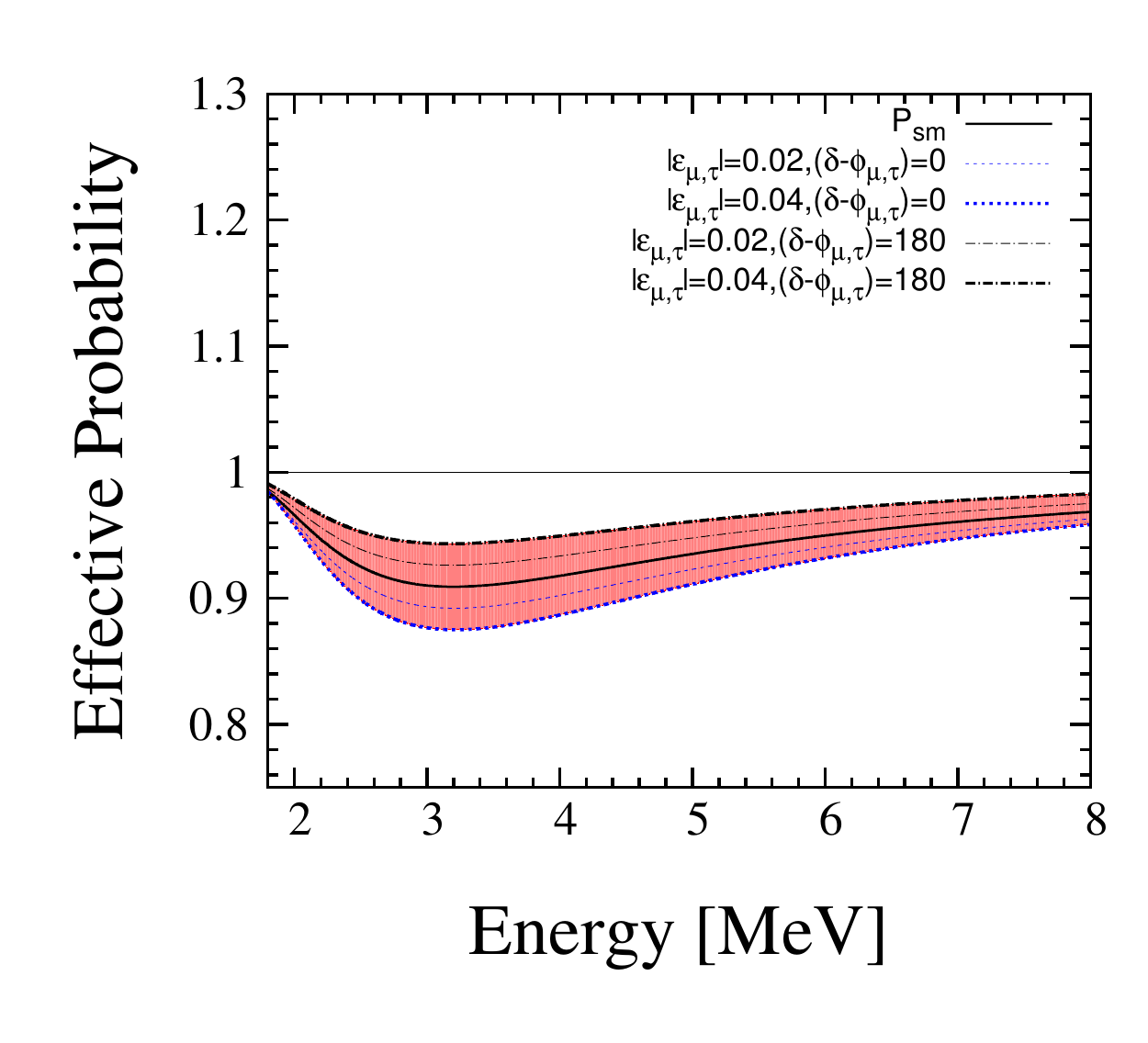}
\caption{Effective $\bar{\nu}^s_e \to \bar{\nu}^d_e$ survival probability as a function 
of neutrino energy in the presence of NSI with $L$ = 1.58 km. The band in the left panel 
has been generated by varying $|\eps_e|$ in the range [0, 0.04] and $\phi_e$ over the 
range [-180$^\circ$, 180$^\circ$] simultaneously. The simultaneous variation of $|\eps_{\mu,\tau}|$ in the 
range [0, 0.04] and ($\delta - \phi_{\mu,\tau}$) over the range [-180$^\circ$, 180$^\circ$] is responsible 
for the band in the right panel. In both the panels, the solid black lines depict the probability without 
new physics involved (SM case).}
\label{fig:prob-band-1}
\end{figure}

Now we will study the possible impact of the NSI parameters at the
effective probability level.  Table~\ref{tab_osc_param_input} depicts
the benchmark values of the various oscillation parameters that are
considered to generate the oscillation probability plots. These
choices of the oscillation parameters are in close agreement with the
best-fit values that have been obtained in the recent global fits of
the world neutrino oscillation
data~\cite{Forero:2014bxa,Capozzi:2013csa,Gonzalez-Garcia:2014bfa}. Here
we would like to mention that for the 2-3 mixing angle, we have taken
the maximal value \ie~$\sa = 0.5$, though in the global fit studies,
there is a slight hint for a non-maximal value of $\tz$. We have also
assumed normal mass ordering, \ie,~$\Delta m_{31}^2$ positive. In
Fig.~\ref{fig:prob-band-1}, we present the standard and the
NSI-modified three-flavor oscillation effective probability as a
function of the electron antineutrino energy with a source-detector
distance of 1.58 km. In the left panel of Fig.~\ref{fig:prob-band-1},
the band shows how the probability changes if we vary $|\eps_e|$ in
the range [0, 0.04] and $\phi_e$ over the range [-180$^\circ$,
180$^\circ$] simultaneously. For the NSI parameters which are only
associated with $\anue$, the probability is independent of the CP
phase $\delta$ (see Eq.~(\ref{eq:pee_sm}) and
Eq.~(\ref{eq:only-nuebar})).  The solid black line shows the standard
oscillation probability without the NSI terms (see
Eq.~(\ref{eq:pee_sm})).  The other four lines have been drawn
considering particular choices of $|\eps_e|$ and $\phi_e$ which are
mentioned in the figure legends. If $\phi_e$ = 180$^\circ$, then the
oscillation probability is less compared to the standard value because
the contribution from the non-oscillatory NSI terms takes the form
$-4|\eps_e|+6|\eps_e|^2$ which always gives an overall negative
contribution to the full probability if $|\eps_e| \leq 0.66$.  On the
other hand, if we consider $\phi_e = 0^\circ$, then the oscillation
probability is above the standard value because the contribution from
the non-oscillatory terms takes the form $4|\eps_e|+6|\eps_e|^2$ which
always gives an overall positive contribution to the full probability
for any choice of $|\eps_e|$.  We can also see that even for a small
value of $|\eps_e|$ of $0.02$, the effective oscillation probability
can be more than unity for most of the energies of interest. This is
the sign of the non-unitarity
effects~\cite{Antusch:2006vwa,FernandezMartinez:2007ms,Goswami:2008mi,Luo:2008vp,Forero:2011pc},
caused by the presence of neutrino NSI at the source and detector of
reactor experiments.  In the right panel of
Fig.~\ref{fig:prob-band-1}, the band shows the changes in the
effective probability after varying $|\eps_{\mu,\tau}|$ in the range
[0, 0.04] and ($\delta - \phi_{\mu,\tau}$) over the range
[-180$^\circ$, 180$^\circ$] simultaneously.  Note that in
Eq.~(\ref{eq:only-numubar}), the phases appear in the form of cosine
of ($\delta - \phi_{\mu,\tau}$) and also the NSI terms have a
non-trivial $L/E$ dependency. The standard oscillation probability
without the NSI terms is shown by the solid black line and the other
four lines have been drawn considering particular choices of
$|\eps_{\mu,\tau}|$ and ($\delta - \phi_{\mu,\tau}$) which are
mentioned in the figure legends.  If ($\delta - \phi_{\mu,\tau}$) =
0$^\circ$ (180$^\circ$), then the effective oscillation probability is
less (more) compared to the standard value for almost all the choices
of neutrino energy as opposed to the case of the NSI parameters
associated with $\anue$.

\begin{figure}[!t]
\centering
\includegraphics[width=0.75\textwidth]{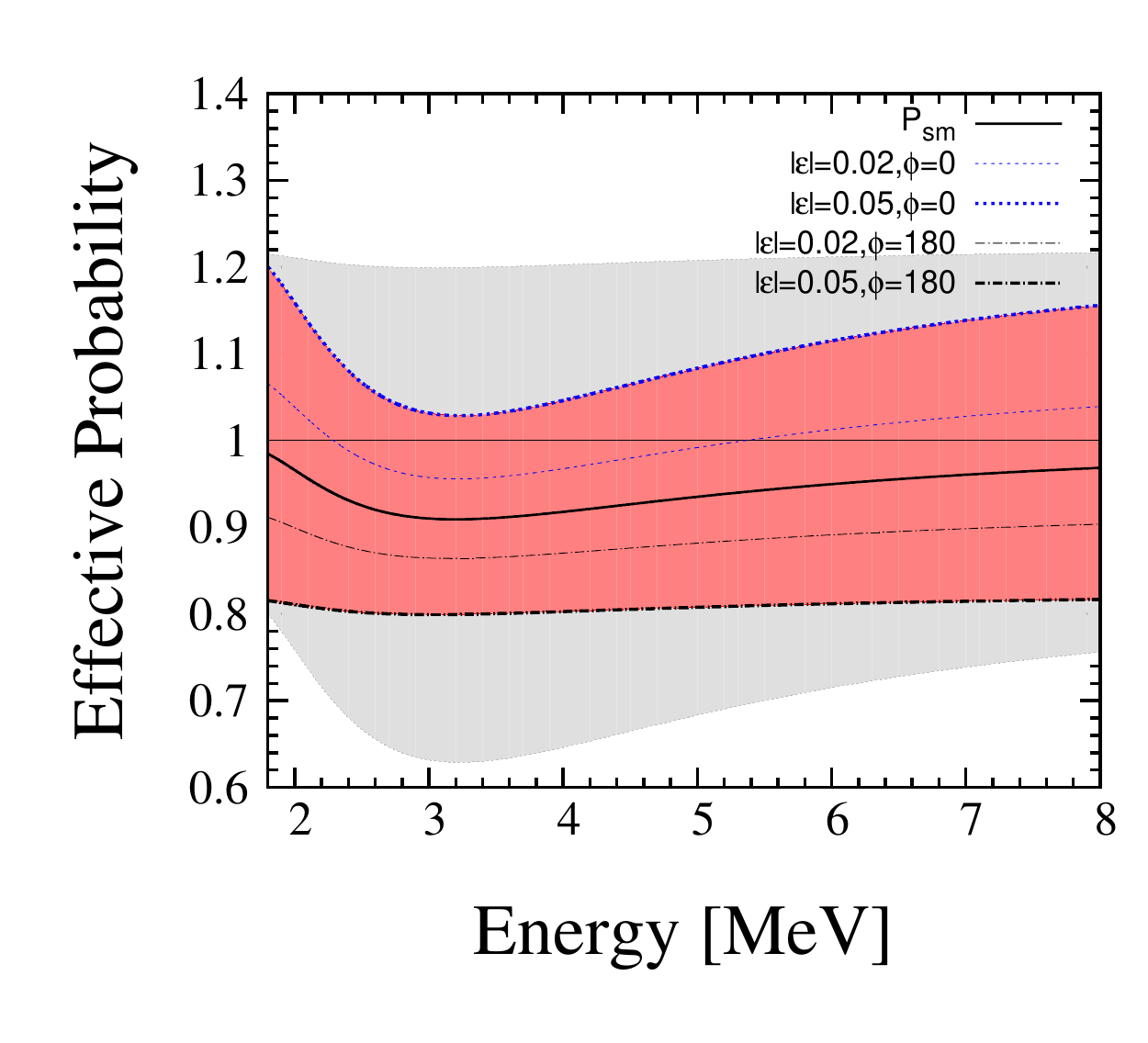}
\caption{Effective $\bar{\nu}^s_e \to \bar{\nu}^d_e$ survival probability as a function of neutrino energy 
with $L$ = 1.58 km for the flavor-universal NSI case (see Sec.~\ref{Special-NSI-Case} for details). 
The dark salmon region shows the combined effect of the variation of the new physics parameters $|\eps|$ 
and $\phi$  with $\delta$ = 0$^\circ$. The extended probability band in light grey 
has been obtained by varying the CP phase $\delta$ in the range [-180$^\circ$, 180$^\circ$] along with the other two 
parameters $|\eps|$ and $\phi$. The solid black line displays the probability without new physics involved (SM case).}
\label{fig:prob-band-2}
\end{figure}

Fig.~\ref{fig:prob-band-2} shows the impact of the NSI parameters at
the probability level with $L$ = 1.58 km for the flavor-universal NSI
case where we consider $|\eps_e| = |\eps_\mu| = |\eps_\tau| = |\eps|$
and $\phi_e = \phi_\mu = \phi_\tau = \phi$.  In this plot, the dark
salmon region has been generated by varying the NSI parameter $|\eps|$
in the range [0, 0.05] and $\phi$ in the range [-180$^\circ$,
180$^\circ$] simultaneously, keeping the CP phase $\delta$ fixed to
0$^\circ$.  Next, we vary $\delta$ in its entire range from
-180$^\circ$ to 180$^\circ$ along with the NSI parameters $|\eps|$ and
$\phi$ and obtain the extended probability band in the form of the
light grey region. In Fig.~\ref{fig:prob-band-2}, the solid black line
depicts the standard probability without considering the NSI
parameters. The other four lines in this plot display the effective
oscillation probability for particular combinations of $|\eps|$ and
$\phi$ with $\delta$ = 0$^\circ$ which are mentioned in the figure
legends. It is quite clear from Eq.~(\ref{eq:flavorless}) that the
non-oscillatory terms dominate over the oscillatory terms in the
flavor-universal NSI case. Therefore, the dark salmon region of
Fig.~\ref{fig:prob-band-2} closely resembles the left panel of
Fig.~\ref{fig:prob-band-1} where we consider the NSI parameters which
are only associated with $\anue$, namely $|\eps_e|$ and
$\phi_e$. Next, we discuss the possible correlations between the NSI
parameters and $\theta_{13}$ with the help of iso-probability plots.

\subsection{Correlations between NSI parameters and $\ty$: iso-probability plots}
\label{iso-probability}

 \begin{figure}[!t]
 \centering
 \includegraphics[width=0.49\textwidth]{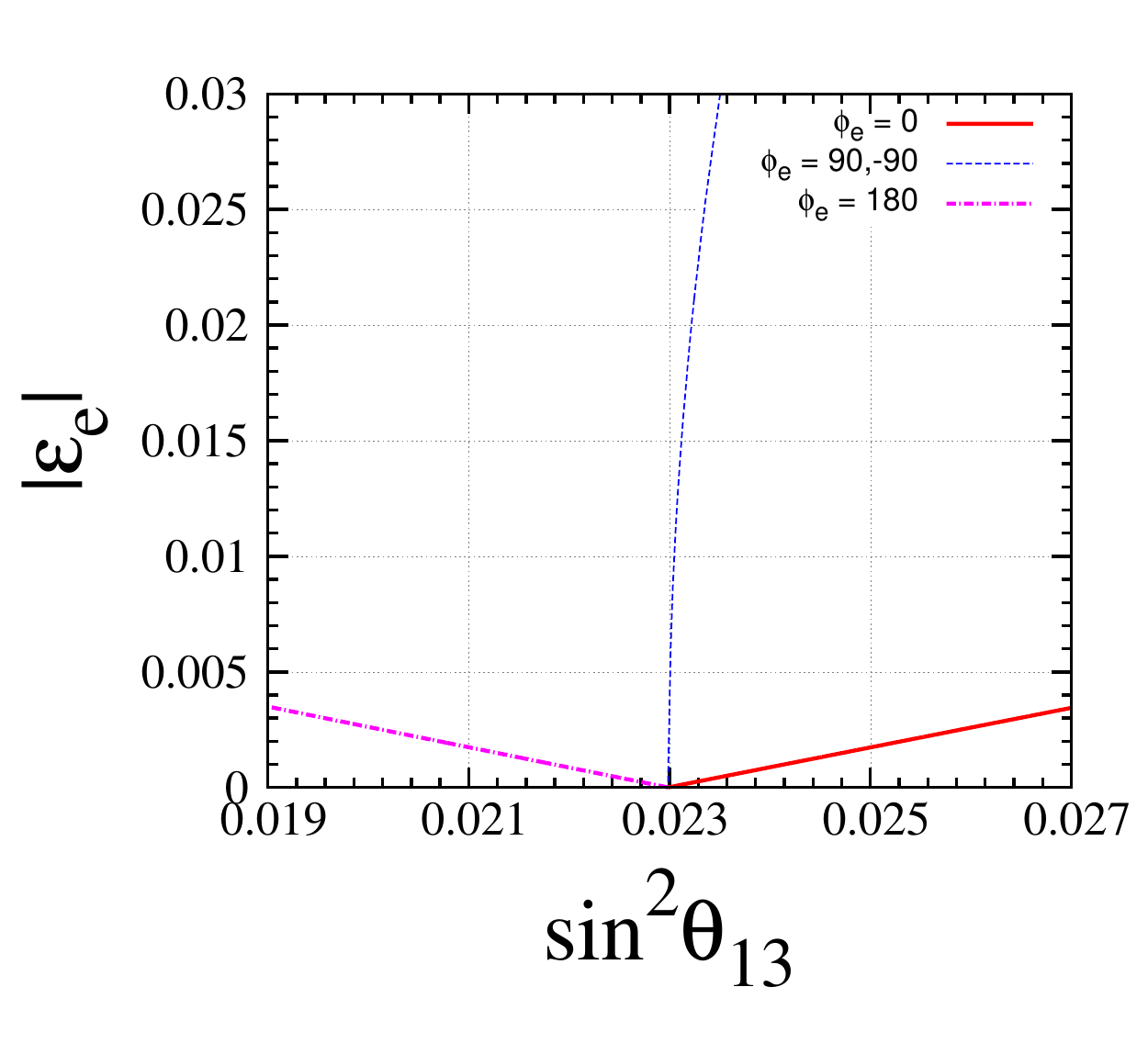}
 \includegraphics[width=0.49\textwidth]{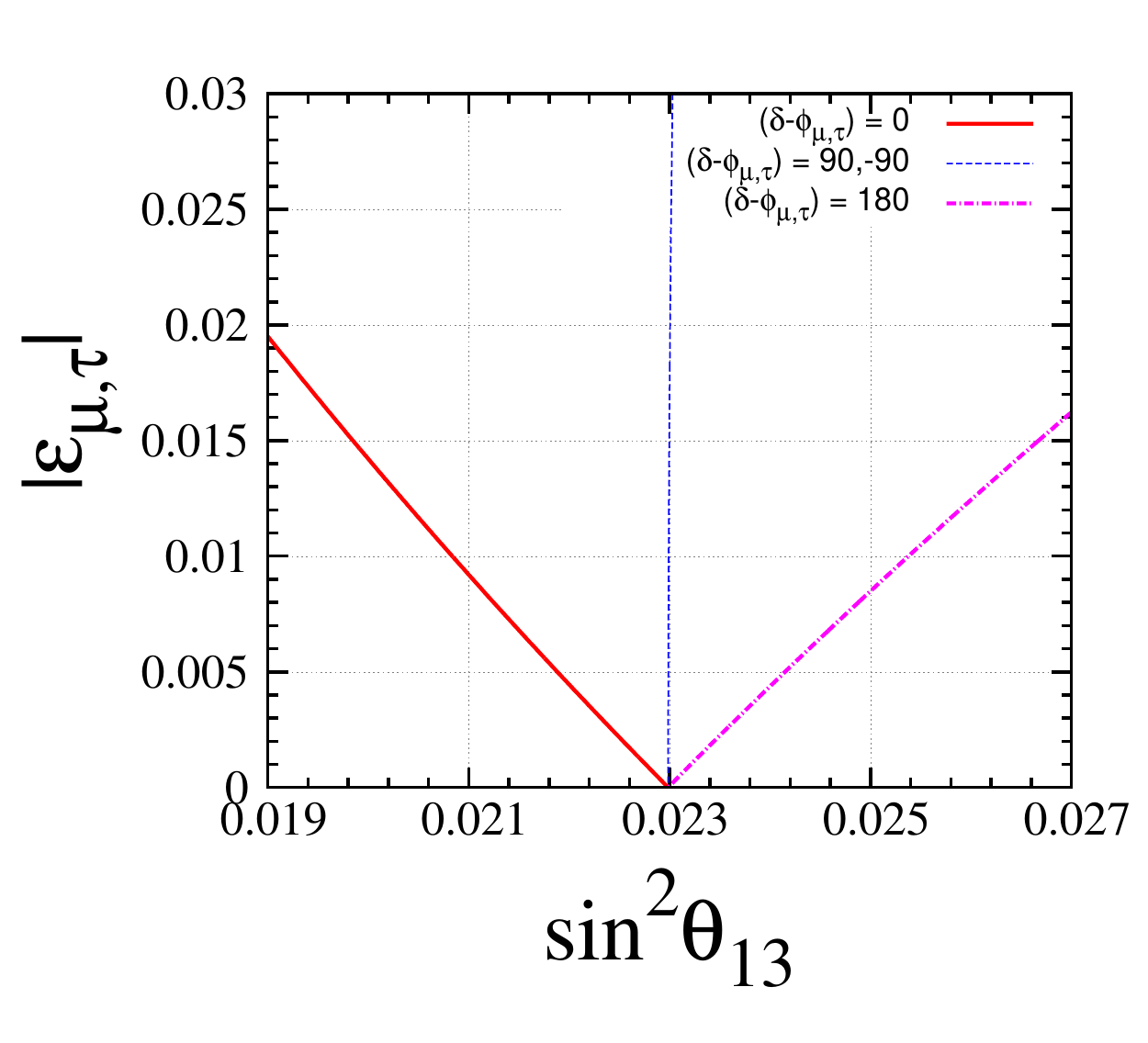}
 \caption{Left panel shows the iso-probability surface contours in the ($\sch$ -- $|\eps_e|$) plane
 for different choices of $\phi_e$ as mentioned in the figure legends. Here we consider $\sch$ = 0.023 and 
 $|\eps_e|$ = 0 as benchmark choices. Right panel displays the same in the ($\sch$ -- $|\eps_{\mu,\tau}|$) plane
 for different choices of ($\delta - \phi_{\mu,\tau}$) considering $\sch$ = 0.023 and $|\eps_{\mu,\tau}|$ = 0
 as true choices. For both the panels, we consider a fixed neutrino energy $E$ = 4 MeV and the baseline
 $L$ = 1.58 km.} 
 \label{fig:isoprob-1}
 \end{figure}

 We consider the neutrino energy $E$ = 4 MeV and the source-detector
 distance $L$ = 1.58 km to draw the iso-probability surface
 plots. Left panel of Fig.~\ref{fig:isoprob-1} shows the
 iso-probability surface contours in the ($\sch$ -- $|\eps_e|$) plane
 for four different choices of $\phi_e$ considering $\sch$ = 0.023 and
 $|\eps_e|$ = 0 as best-fit choices.  Our best-fit choices correspond
 to the standard oscillation probability without considering the NSI
 parameters as given by Eq.~(\ref{eq:pee_sm}). Now as we consider the
 finite value of $|\eps_e|$ with $\phi_e$ = 0$^\circ$ (see the solid
 red line), the NSI terms increase the overall probability (see
 Eq.~(\ref{eq:only-nuebar})).  Then, we need to increase the value of
 $\sch$ to reduce the SM probability in order to compensate the
 enhancement due to the NSI contribution. For $\phi_e$ = 180$^\circ$
 case (see the dot-dashed magenta line), a finite value of $|\eps_e|$
 decreases the overall probability demanding a lower value of $\sch$
 so as to enhance the contribution from the standard probability. In
 the cases with $\phi_e$ = 90$^\circ$ or -90$^\circ$ (see the dashed
 blue line), the non-oscillatory NSI term which is linear in
 $|\eps_e|$ drops out from the probability expression and the
 remaining NSI contribution is $2|\eps_e|^2$ (see
 Eq.~(\ref{eq:only-nuebar})). Due to this weak quadratic dependence on
 $|\eps_e|$, a very large value of $|\eps_e|$ is needed to compensate
 even a very small increment in $\sch$. Right panel displays the same
 in the ($\sch$ -- $|\eps_{\mu,\tau}|$) plane for different choices of
 ($\delta - \phi_{\mu,\tau}$). Here $\sch$ = 0.023 and
 $|\eps_{\mu,\tau}|$ = 0 are considered as true choices. In this
 panel, the combination of phases ($\delta - \phi_{\mu,\tau}$) shows
 opposite features for 0$^\circ$ and 180$^\circ$ as compared to
 $\phi_e$ in the left panel. Note that the impact of
 $|\eps_{\mu,\tau}|$ on $\sch$ is weaker compared to $|\eps_e|$.  If
 we examine the terms which are linear in $|\eps_e|$ (see
 Eq.~(\ref{eq:only-nuebar})) and $|\eps_{\mu,\tau}|$ (see
 Eq.~(\ref{eq:only-numubar})) then we can see that the contribution
 coming from $|\eps_{\mu,\tau}|$ is $\sin\theta_{13}$ suppressed even
 if we work at the first oscillation maximum.  For ($\delta -
 \phi_{\mu,\tau}$) = 90$^\circ$ or -90$^\circ$, the
 $|\eps_{\mu,\tau}|$-dependent terms completely disappear from
 Eq.~(\ref{eq:only-numubar}) if $\sa$ = 0.5 and $\sin^2\Delta_{31}$ =
 1. Therefore, we do not see any correlation between $\sch$ and
 $|\eps_{\mu,\tau}|$ for ($\delta - \phi_{\mu,\tau}$) = 90$^\circ$ or
 -90$^\circ$ in the right panel of Fig.~\ref{fig:isoprob-1}.

\begin{figure}[!t]
 \centering
 \includegraphics[width=0.49\textwidth]{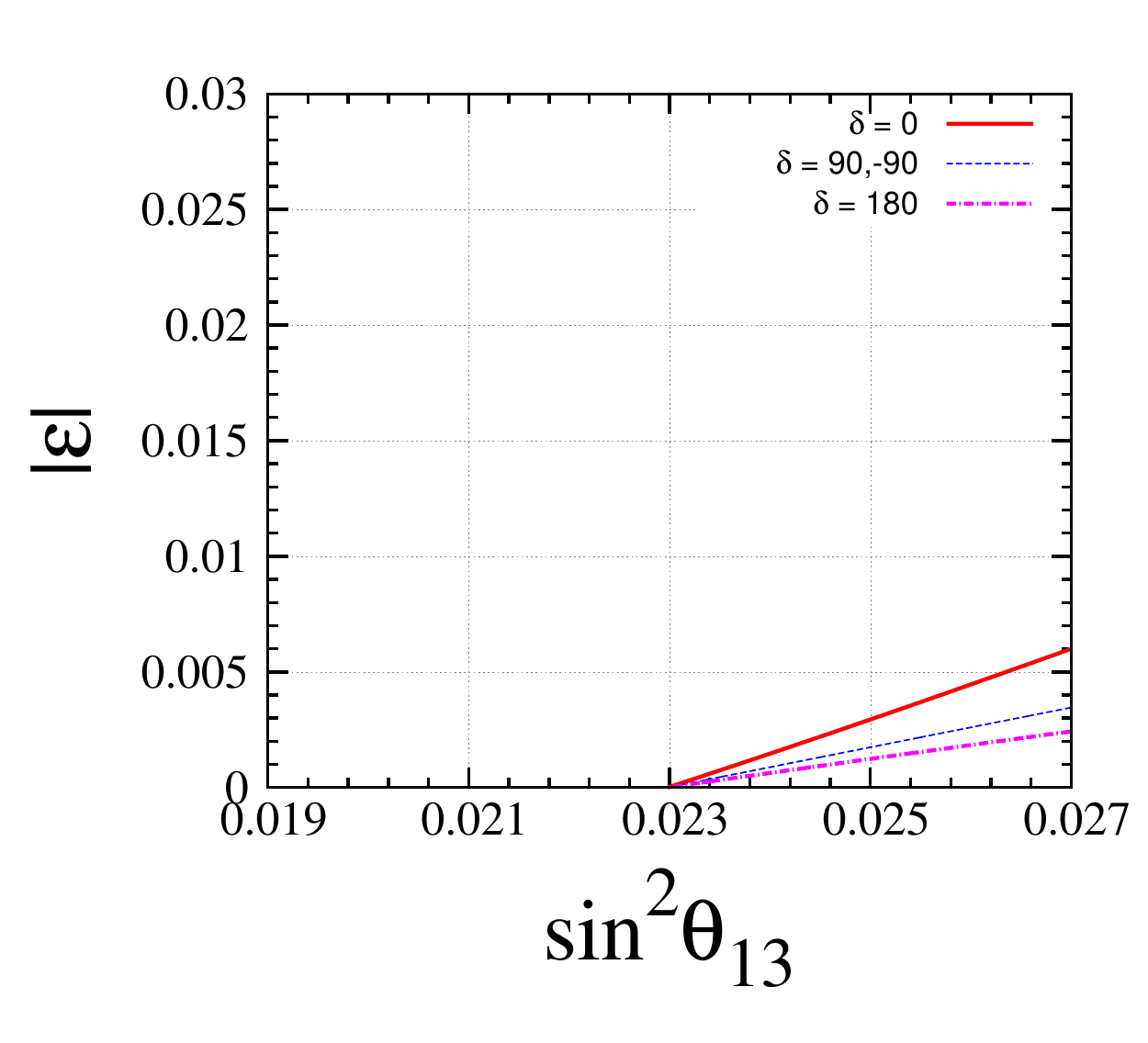}
 \includegraphics[width=0.49\textwidth]{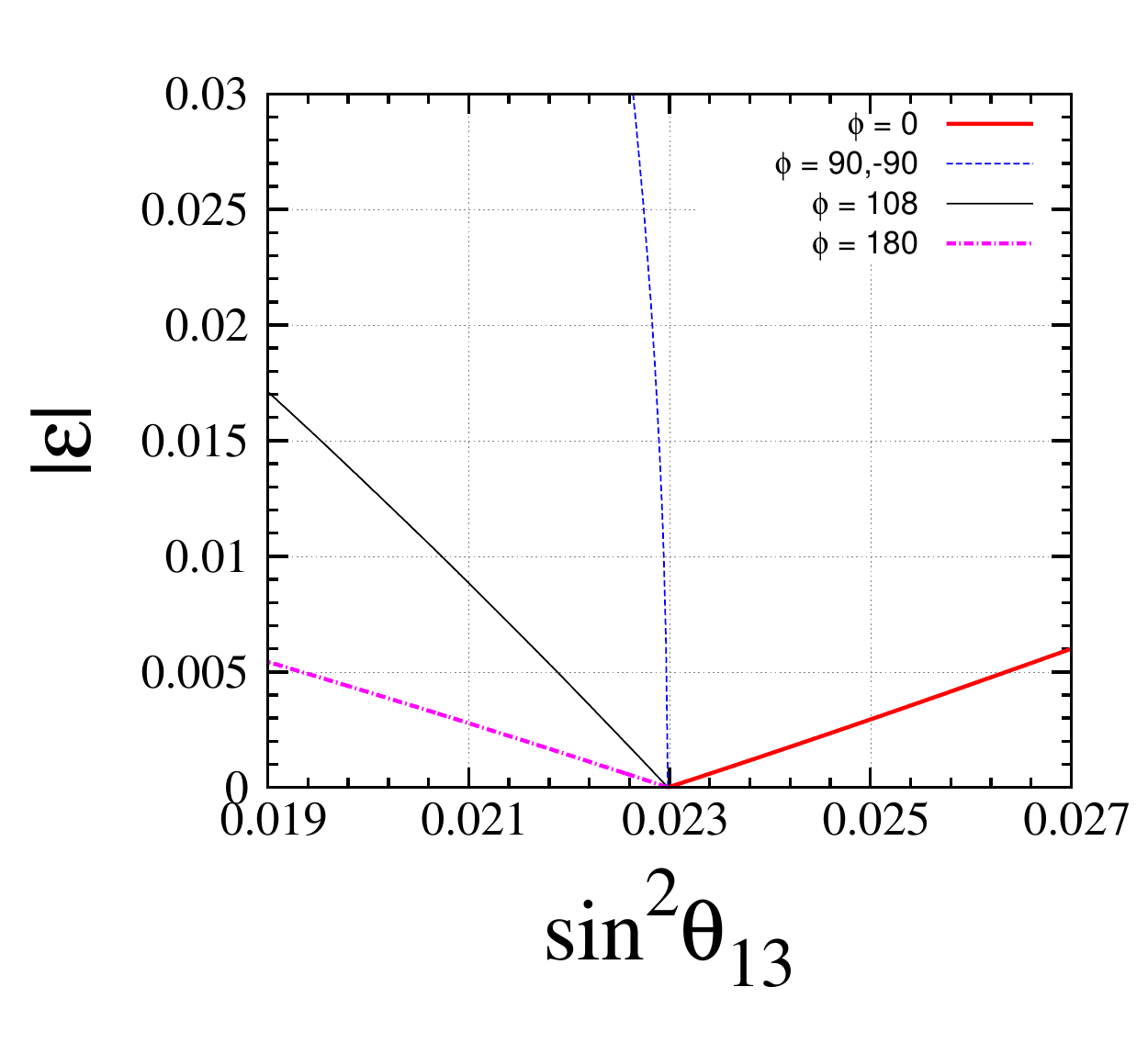}
 \caption{Left panel shows the iso-probability surface contours for the flavor-universal NSI case in the 
 ($\sch$ -- $|\eps|$) plane for different choices of the CP phase $\delta$ with $\phi$ = 0$^\circ$. 
 Here we consider $\sch$ = 0.023 and $|\eps|$ = 0 as benchmark choices. Right panel displays the same 
 for various choices of $\phi$ assuming $\delta$ = 0$^\circ$. For both the panels, we take a fixed neutrino 
 energy $E$ = 4 MeV and the baseline $L$ = 1.58 km.}
\label{fig:isoprob-2} 
 \end{figure}

 In Fig.~\ref{fig:isoprob-2}, we show the correlation between $\sch$
 and $|\eps|$ for the flavor-universal NSI case with the help of
 iso-probability surface contours. In the left panel, we consider four
 different choices of the CP phase $\delta$ keeping $\phi$ fixed to
 0$^\circ$. In the right panel, we take five different choices of
 $\phi$ assuming $\delta$ = 0$^\circ$. In both panels, the
 iso-probability surface contours are the same for 90$^\circ$ and
 -90$^\circ$ choices of phases (see the dashed blue lines) because the
 phases appear in the form of cosines in
 Eq.~(\ref{eq:flavorless}). For the $\phi$ = 0$^\circ$ case (left
 panel), the contribution from the non-oscillatory NSI terms is
 maximum and takes the form $4|\eps|+10|\eps|^2$. Thus, it enhances
 the overall oscillation probability to a great extent even for a
 small value of $|\eps|$. Now to compensate this enhancement, we need
 to increase the value of $\sch$ to reduce the contribution coming
 from the standard survival oscillation probability.  Right panel of
 Fig.~\ref{fig:isoprob-2} ($\delta$ = 0$^\circ$ case) closely
 resembles the left panel of Fig.~\ref{fig:isoprob-1} because for the
 flavor-universal NSI case, the contribution from the non-oscillatory
 NSI terms dominates which is also true for the NSI parameters
 associated with $\anue$ and also these NSI parameters are $\delta$
 independent.
 In the right panel, we present a special case of $\phi$ = 108$^\circ$
 to explain the features emerging from the extreme right panel of
 Fig.~\ref{fig:eps_univ} (see later in Sec.~\ref{NSI-bounds}) where we
 have displayed the allowed region in ($\sch$ -- $|\eps|$) plane using
 the current data from the Daya Bay experiment allowing the
 flavor-universal NSI phase $\phi$ to vary in the entire range of
 [-180$^\circ$, 180$^\circ$] with the CP phase $\delta$ to be fixed to
 0$^\circ$.
 Eq.~(\ref{eq:flavorless}) suggests that we can always choose some
 values of $\phi$ such that the non-oscillatory terms are
 canceled. This happens, for instance, for $\phi = 108^\circ$ and
 $|\eps| = 0.206$. In this case, the dominant oscillatory NSI terms
 give a negative contribution to the neutrino survival probability
 that, to be compensated, requires an enhancement of the standard
 survival probability by reducing the value of $\sin^2 \theta_{13}$ by
 the same quantity.
This is exactly the feature that we can see for $\phi = 108^\circ$ case.

\section{Data analysis of modern reactor experiments}
\label{Analysis-DayaBay}

Reactor antineutrinos are produced by the fission of the isotopes
$^{235}$U, $^{239}$Pu, $^{241}$Pu and $^{238}$U contributing to the
neutrino flux with a certain fission fraction $f_k$.
Reactor antineutrinos are detected via inverse $\beta$-decay process
(IBD), $\bar{\nu_e}+p \to e^+ + n$. The technique used is a delayed
coincidence between two gamma rays: one coming from the positron
(\textit{prompt} signal) and the other coming from the neutron capture
in the innermost part of the antineutrino detector (AD), containing
gadolinium-doped liquid scintillator. The light created is collected
by the photo-multipliers (PMTs) located in the outermost mineral
oil-region. The antineutrino energy $E_{\bar{\nu}}$ is reconstructed
from the positron prompt energy $E_{\text{prompt}}$ following the
relation: $E_{\bar{\nu}}=E_{\text{prompt}}+\bar{E}_n+0.78\,
\text{MeV}$, where $\bar{E}_n$ is the
average neutron recoil energy. \\
The expected number of IBD events at the $d$-th detector, $T_d$, can
be estimated summing up the contributions of all reactors to the
detector:
\begin{eqnarray}\label{eq:th_osc}
T_d & = &  \sum_r T_{rd} = \nonumber \\
 & = & \sum_r \epsilon_d \frac{N_p}{4\pi L_{rd}^2} \frac{P_{th}^r}{\sum_k f_k 
\langle E_k \rangle} \sum_k f_k \int_0^{\infty} dE \, \Phi_k(E) 
\,\sigma_{IBD}(E) P_{ee}(E,L_{rd}),
\end{eqnarray}
where $N_p$ is the number of protons in the target volume, $P_{th}^r$
is the reactor thermal power, $\epsilon_d$ denotes the efficiency of
the detector and $\langle E_k \rangle$ is the energy release per
fission for a given isotope $k$ taken from
Ref.~\cite{Kopeikin:2004cn}. The neutrino survival probability
$P_{ee}$ depends also on the distance from $r$-th reactor to $d$-th
detector, $L_{rd}$.
For the antineutrino flux prediction $\Phi_k(E)$ we use the
parameterization given in Ref.~\cite{Mueller:2011nm} as well as the
new normalization for reactor antineutrino fluxes updated in
Ref.~\cite{Abazajian:2012ys}.
The inverse beta decay cross section $\sigma_{\text{IBD}}(E_\nu)$ is taken from
Ref.~\cite{Vogel:1999zy}. \\
%

\subsection*{Daya Bay Experiment}

Daya Bay is a reactor neutrino experiment with several antineutrino
detectors (ADs), arranged in three experimental halls (EHs).  Electron
antineutrinos are generated in six reactor cores, distributed in
pairs, with equal thermal power (P$_{th}^r$=2.9 GW$_{th}$) and
detected in the EHs. The effective baselines are 512 m and 561 m for
the near halls EH1 and EH2 and 1579 m for the far hall
EH3~\cite{An:2013zwz}. With this near-far technology Daya Bay has
minimized the systematic errors coming from the ADs and thus provided
until now the most precise determination of the reactor mixing angle.
In the last Neutrino conference, Daya Bay has reported its preliminary results 
considering 621 days of data taking combining their results for two different experimental
setups~\cite{talk-Zhang-Nu2014}: one with six ADs as it was published
in Ref.~\cite{An:2013zwz} and the other after the installation of two
more detectors, eight ADs in total.
This new combined data set has four times more statistics in
comparison with the previous Daya Bay results. Thus, the precision in
the determination of the reactor mixing angle has been improved, and
it is now of the order of 6$\%$.

In this work we will consider the most recent data release by the Daya
Bay Collaboration described above and we will concentrate on the total
observed rates at each detector, that will be analyzed using the
following $\chi^2$ expression:
\begin{eqnarray}  
\label{eq:chi2}
 \chi^2 & = &  \sum_{d=1}^{8}
 \frac{\left[M_d-T_d\left(1+  a_{\rm norm}
 + \sum_r\omega_r^d\alpha_r
 + \xi_d\right) +\beta_d\right]^2}
 {M_d+B_d} \nonumber \\
 & + &
 \sum_{r=1}^{6}\frac{\alpha_r^2}{\sigma_r^2}
 + \sum_{d=1}^{8} \left(
 \frac{\xi_d^2}{\sigma_d^2}
 + \frac{\beta_d^2}{\sigma_{B}^2}\right) +
 \frac{a^2_{\rm norm}}{\sigma^2_a}.
\end{eqnarray}
Here $T_d$ corresponds to the theoretical prediction in
Eq. (\ref{eq:th_osc}), $M_d$ is the measured number of events at the
$d$-th AD with its backgrounds ($B_d$) subtracted and $\omega_r^d$ is
the fractional contribution of the $r$-th reactor to the $d$-th AD
number of events, determined by the baselines $L_{rd}$ and the total
thermal power of each reactor.
The pull parameters, used to include the systematical errors in the
analysis, are given by the set $(\alpha_r,\xi_d,\beta_d)$ representing
the reactor, detector and background uncertainties with the
corresponding set of errors
$(\sigma_r,\sigma_d,\sigma_B)$. Uncertainties in the reactor related
quantities are included in $\sigma_r$ ($0.8\%$) while the uncorrelated
combined uncertainties in the ADs are included in $\sigma_d$
($0.2\%$). $\sigma_B$ is the quadratic sum of the background
uncertainties taken from Ref.~\cite{talk-Zhang-Nu2014}.
Finally, we also consider an absolute normalization factor $a_{\rm
  norm}$ to account for the uncertainty in the total normalization of
events at the ADs, given by $\sigma_a$, and coming for instance from
uncertainties in the normalization of reactor antineutrino fluxes.
In our analysis we will follow two different approaches concerning
this parameter. In Sec.~\ref{NSI-bounds} we will take it equal to
zero, assuming perfect knowledge of the events normalization. This
hypothesis will be relaxed in Sec.~\ref{NSI-bounds-freenorm}, where we
will allow for a non-zero normalization factor in the statistical
analysis, being determined from the fit to the Daya Bay data. As we
will see, the results obtained in our analysis are strongly correlated
with the treatment of the total normalization of reactor neutrino
events in the statistical analysis of Daya Bay data and therefore it
is of crucial importance to do a proper treatment of this factor.

\section{Bounds on NSI from  Daya Bay without normalization error}
\label{NSI-bounds}

In this section we will present the bounds on the NSI couplings we
have obtained using current Daya Bay reactor data.
In all the results, we have assumed maximal 2-3 mixing and we have
marginalized over atmospheric splitting with a prior of 3\%.
For definiteness we will start considering only the couplings relative
to electron neutrino: ($|\eps_e|$,$\phi_e$), for what we will switch
all the other NSI parameters to zero. Next we will do the same for
($|\eps_\mu|$,$\phi_\mu$) and ($|\eps_\tau|$,$\phi_\tau$), that are
equivalent for maximal value of $\theta_{23}$. Finally, we will
consider the possibility of having all NSI couplings with the same
value: $\eps_e = \eps_\mu = \eps_\tau = \eps$. In all cases we will
discuss the bounds arising from Daya Bay data in comparison with
existing bounds. We will also consider the robustness of the
$\theta_{13}$ measurement by Daya Bay in the presence of NSI.

\subsection{Constraints on electron-NSI couplings}
\label{subsec:nsie}

\begin{figure}[!tb]
\centerline{
\includegraphics[width=0.49\textwidth]{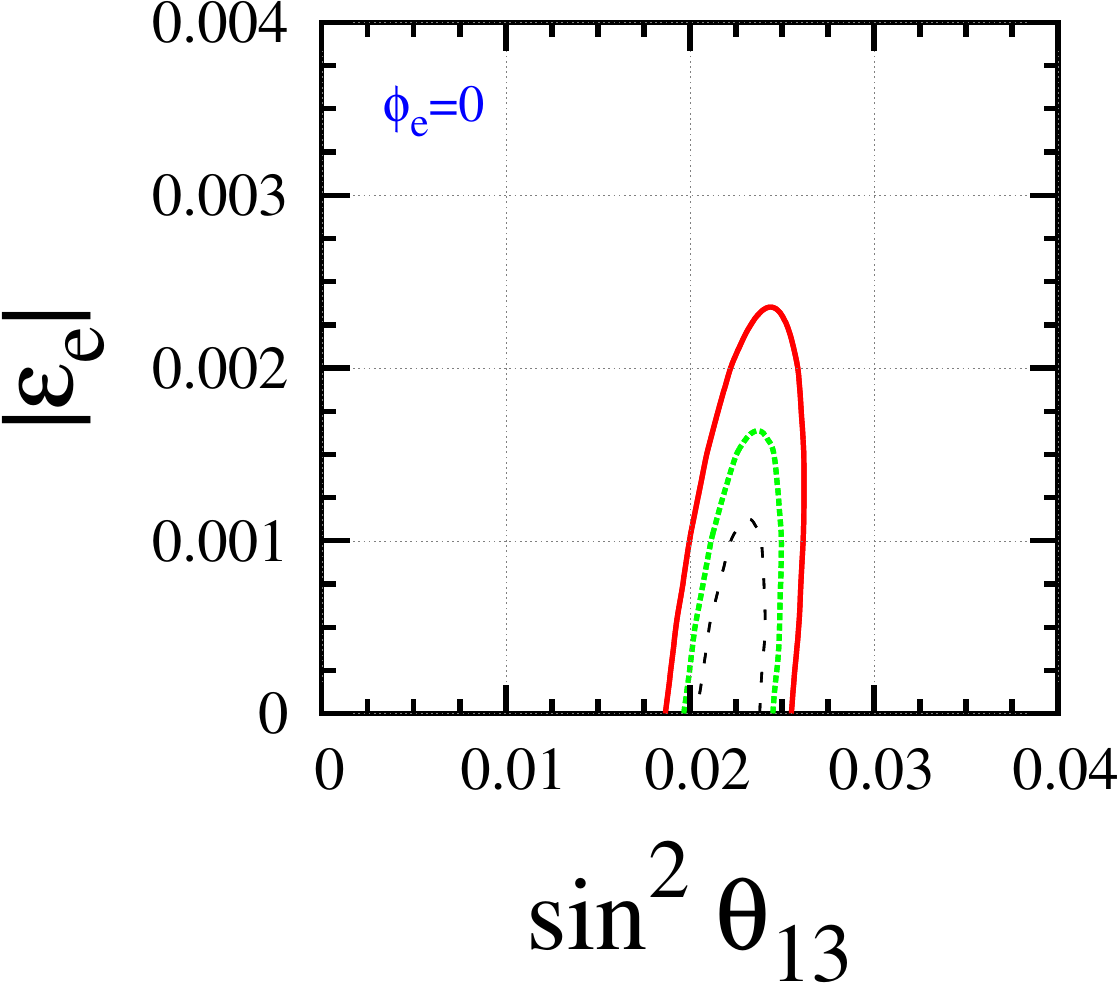}
\includegraphics[width=0.49\textwidth]{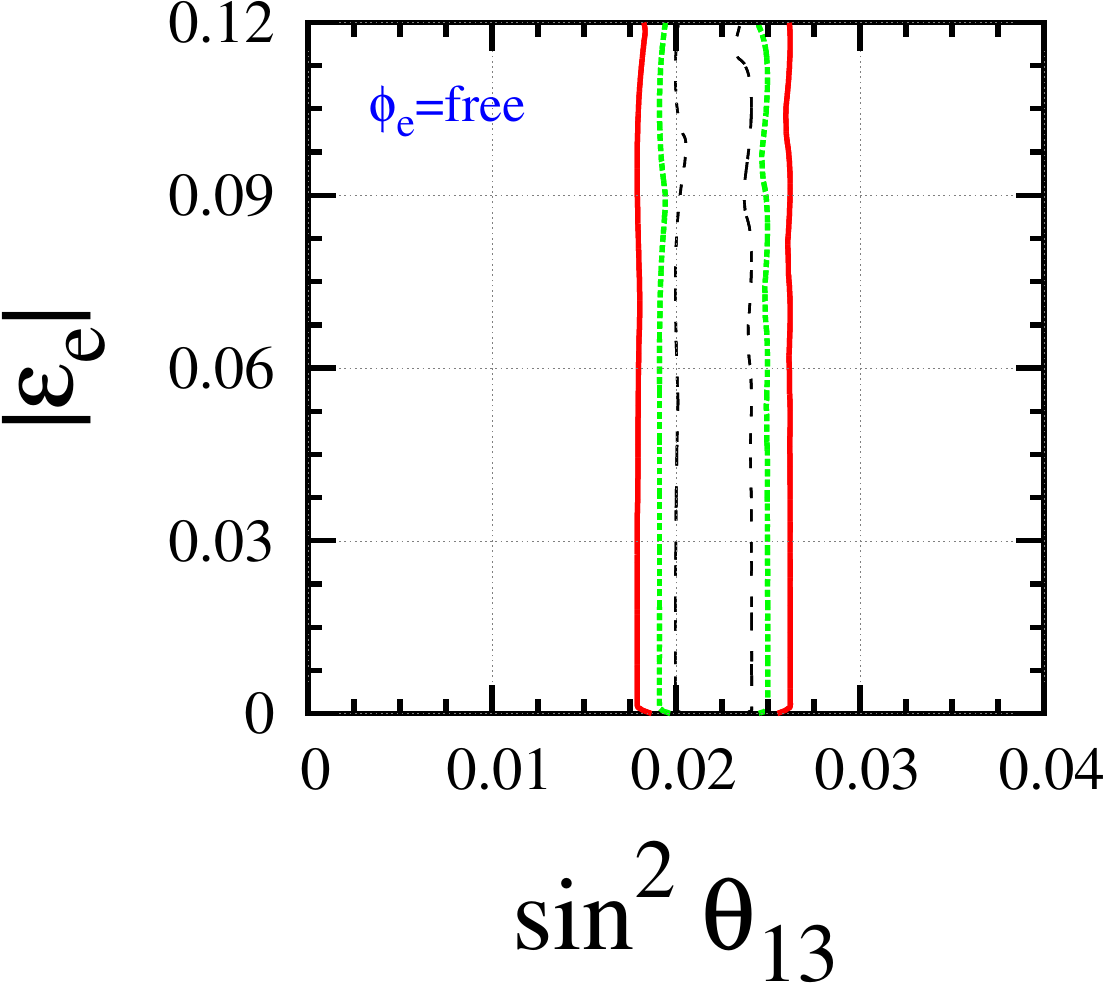}
}
\caption{Allowed region in the $\sin^2\theta_{13}$ - $|\eps_e|$ plane. Left panel is obtained by setting the phase $\phi_e$ to zero, while in the right panel $\phi_e$ is marginalized, varying freely between -$180^\circ$ and $180^\circ$. The regions correspond to 68\% (black dashed line), 90\%  (green line) and 99\% C.L. (red line) for 2 d.o.f. }
\label{fig:eps_e}
\end{figure}

According to the expression in Eq.~(\ref{eq:only-nuebar}), the
effective survival probability in the case when only NSI with electron
antineutrinos are considered is independent of the standard CP phase
$\delta$. Therefore, in our analysis we consider only two cases, one
with the only relevant phase $\phi_e$ set to zero, and a second case
where we allow this phase to vary freely. Our results are presented in
Fig.~\ref{fig:eps_e}.
%
%
From the left panel in this figure, we can confirm the behaviors shown
by the iso-probability curves in the Sec.~\ref{iso-probability},
namely, the presence of a non-zero $\eps_e$ coupling has to be
compensated with a slightly larger value of the reactor mixing angle
$\ty$.
In this panel one also sees how current Daya Bay data constrain very
strongly the magnitude of the NSI coupling $|\eps_e|$, improving the
current bound in Eq.~(\ref{eq:bound-biggio}) by one order of
magnitude:
\begin{equation}
|\eps_e| \le 1.2\times 10^{-3}  \quad \, (90\% \, \text{C.L.})\,.
\label{eq:bound_eps_e}
\end{equation}
However, the situation changes dramatically when the phase $\phi_e$ is
allowed to vary freely, as shown in the right panel of
Fig.~\ref{fig:eps_e}. In this case, the strong bound on $|\eps_e|$
disappears due to the presence of a correlation bewteen  $|\eps_e|$ and $\cos\phi_e$ in the $|\eps_e|$-linear term in the neutrino survival
probability (see Eq.(\ref{eq:only-nuebar})).  
The presence of second order terms in $|\eps_e|$  is not enough to break this degeneracy and, therefore, the sensitivity to $|\eps_e|$
disappears and no bound can be obtained from reactor data.

Concerning the determination of the reactor mixing angle, the presence
of the NSI-coupling with electron antineutrinos results in the
following allowed range for $\theta_{13}$:
\begin{equation}
0.020\le \sin^2\theta_{13}\le 0.024 \quad \, (90\% \, \text{C.L.})\,.
\end{equation}
The same interval is obtained for the two panels at
Fig.~\ref{fig:eps_e} and it also coincides exactly with the allowed
range in absence of NSI. In consequence, we can say that the reactor
angle determination by Daya Bay is robust in this specific case.
%

\subsection{Constraints on muon/tau-NSI couplings}
\label{subsec:nsimu}

In this subsection we present the results obtained considering only
the NSI parameters associated with muon and tau neutrinos. As we have
discussed in Sec.~\ref{Special-NSI-Case}, in this case, the phases
$\delta$ and $\phi_{\mu,\tau}$ do not appear separately in the
expression of the survival probability, see
Eq.~(\ref{eq:only-numubar}). Therefore, it is enough to consider in
our calculations the effective phase ($\delta - \phi_{\mu,\tau}$).

\begin{figure}[!tb]
\centerline{
\includegraphics[width=0.49\textwidth]{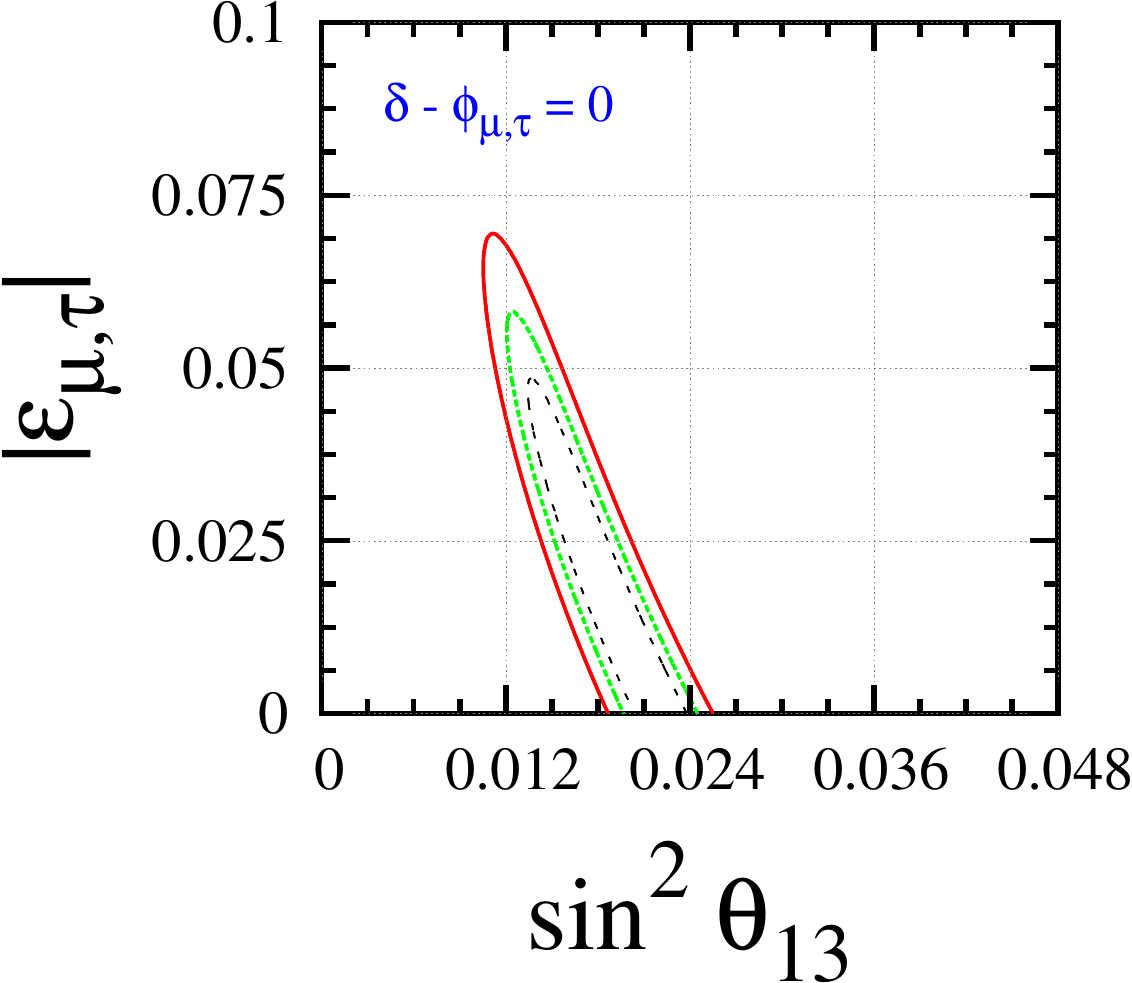}
\includegraphics[width=0.49\textwidth]{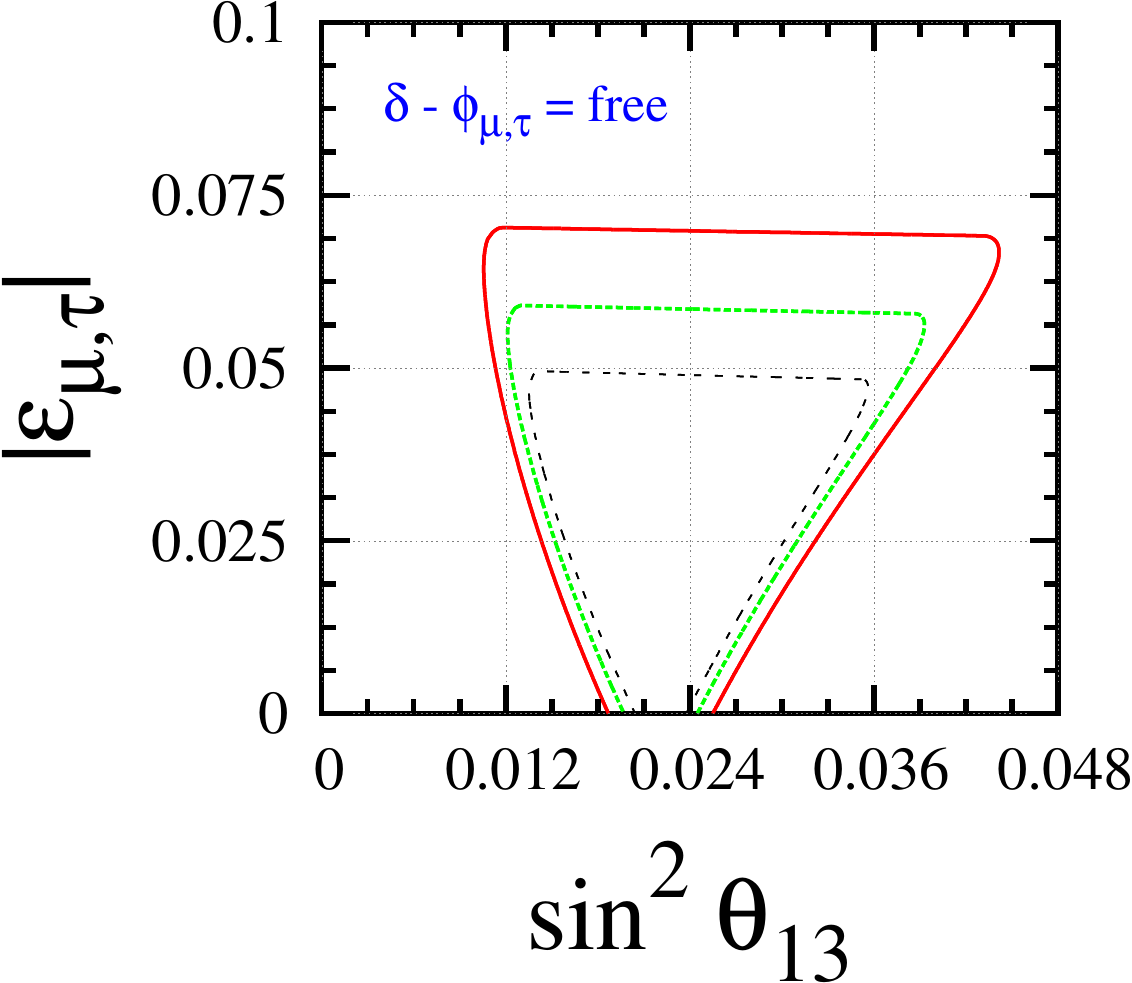}
}
\caption{Allowed region in the $\sin^2\theta_{13}$ - $|\eps_{\mu,\tau}|$ plane. Left panel is obtained setting the relevant phase ($\delta$ - $\phi_{\mu,\tau}$) to zero, while in the right panel ($\delta$ - $\phi_{\mu,\tau}$) is marginalized, varying freely between -$180^\circ$ and $180^\circ$. The regions correspond to 68\% (black dashed line), 90\% (green line), and 99\% C.L. (red line) for 2 d.o.f. }
\label{fig:eps_x}
\end{figure}


The results corresponding to this particular case are shown in
Fig.~\ref{fig:eps_x}.
Here again we can see how the regions presented in the left panel of
the figure agree with the behavior shown in the iso-probability plots
(right panel of Fig.~\ref{fig:isoprob-1}) where there is an
anticorrelation between the reactor angle and the NSI coupling
$|\varepsilon_{\mu,\tau}|$. Thus, an increase in
$|\varepsilon_{\mu,\tau}|$ is compensated by a shift of the preferred
value of the reactor mixing angle toward smaller values, differently
to what happens with the NSI coupling $|\varepsilon_e|$ in
Fig.~\ref{fig:eps_e}.  The allowed interval for $\theta_{13}$ in this
case is given by:
\begin{equation}
0.013\le \sin^2\theta_{13}\le 0.024  \quad \, (90\%  \, \text{C.L.})\, ,
\end{equation}
while the obtained bound for the NSI coupling is the following
\begin{equation}
|\eps_{\mu,\tau}| \le 5.1\times 10^{-2} \quad \, (90\%  \, \text{C.L.})\,.
\end{equation}
In this case reactor data can not improve the present constraints on
the NSI couplings at Eq.~(\ref{eq:bound-biggio}), and we get a limit
of the same order of magnitude of the ones derived at
Ref.~\cite{Biggio:2009nt}.  However, in both cases the limits have
been derived using different data and assumptions, and therefore, they
can be regarded as complementary bounds coming from different data
sets.

In the right panel of Fig.~\ref{fig:eps_x} we show the results
obtained when the phase ($\delta-\phi_{\mu,\tau}$) is allowed to vary.
In this case, a wider range in the reactor mixing angle is allowed:
\begin{equation}
0.013\le \sin^2\theta_{13}\le 0.036  \quad \, (90\%  \, \text{C.L.}) \,.
\end{equation}
The reason is that, in addition to the anticorrelation shown in the
left panel, a correlation between the reactor angle and
$|\varepsilon_{\mu,\tau}|$ is also possible when the cosine function
in Eq.~(\ref{eq:shift-mu}) is negative. 
Note, however, that both correlations are not symmetric, what results in the asymmetric 
behaviour of the allowed $\sin^2\theta_{13}$ region
with a bigger enlargement in the direction of increasing $\theta_{13}$. This can be explained 
by the presence 
of a linear term in $\sin\theta_{13}$ in the redefinition of the 
effective reactor angle in the presence of NSI given in Eq.~(\ref{eq:shift-mu}).
Nevertheless, even though there is a wider allowed region in the reactor angle, the bound
on the $|\varepsilon_{\mu,\tau}|$ NSI coupling is nearly the same as
the one obtained when the phase ($\delta-\phi_{\mu,\tau}$) is set to
zero, namely:
\begin{equation}
|\eps_{\mu,\tau}| \le 5.2\times 10^{-2} \quad \, (90\%  \, \text{C.L.})\,.
\end{equation}
%

\subsection{Constraints for the flavor-universal NSI case}
\label{subsec:FU}

\begin{figure}[!tb]
\centerline{
\includegraphics[width=0.33\textwidth]{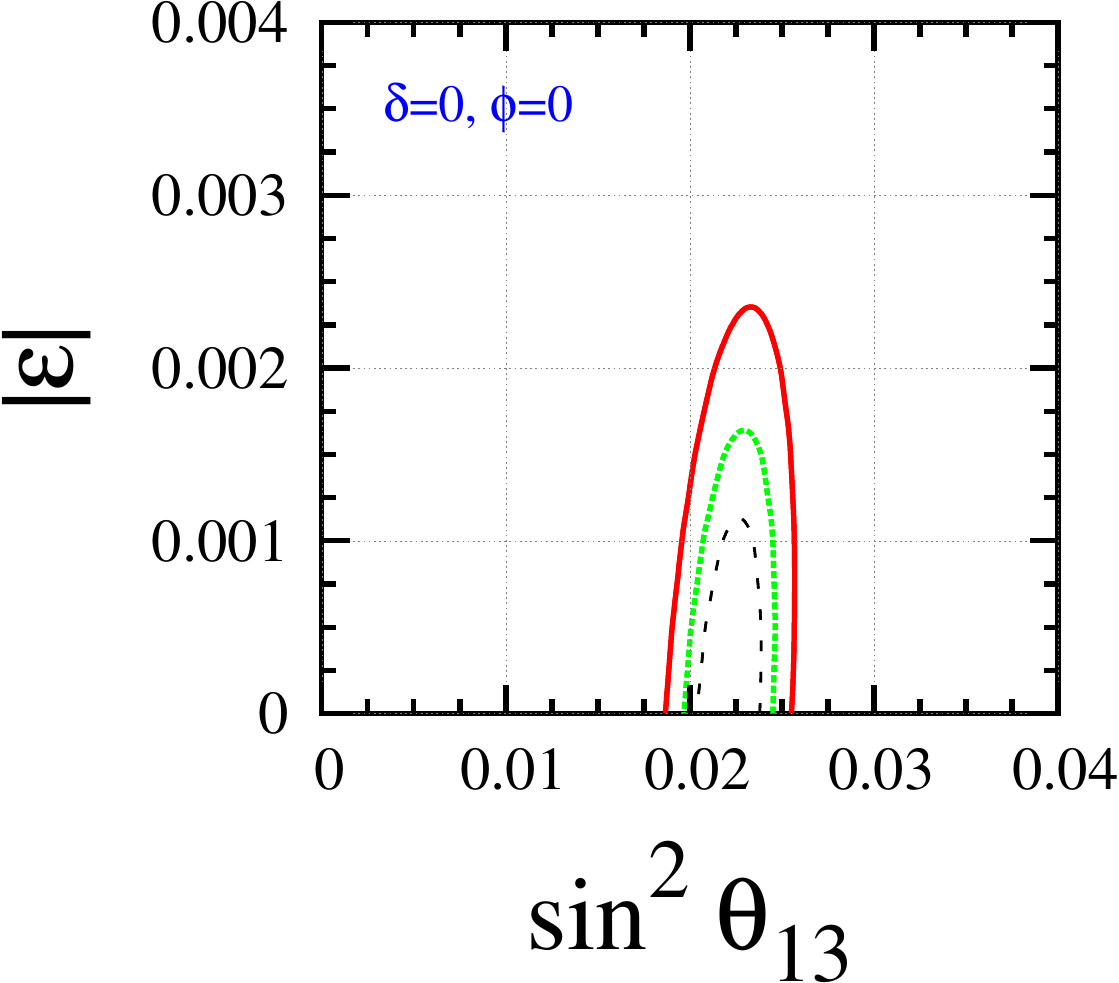}
\includegraphics[width=0.33\textwidth]{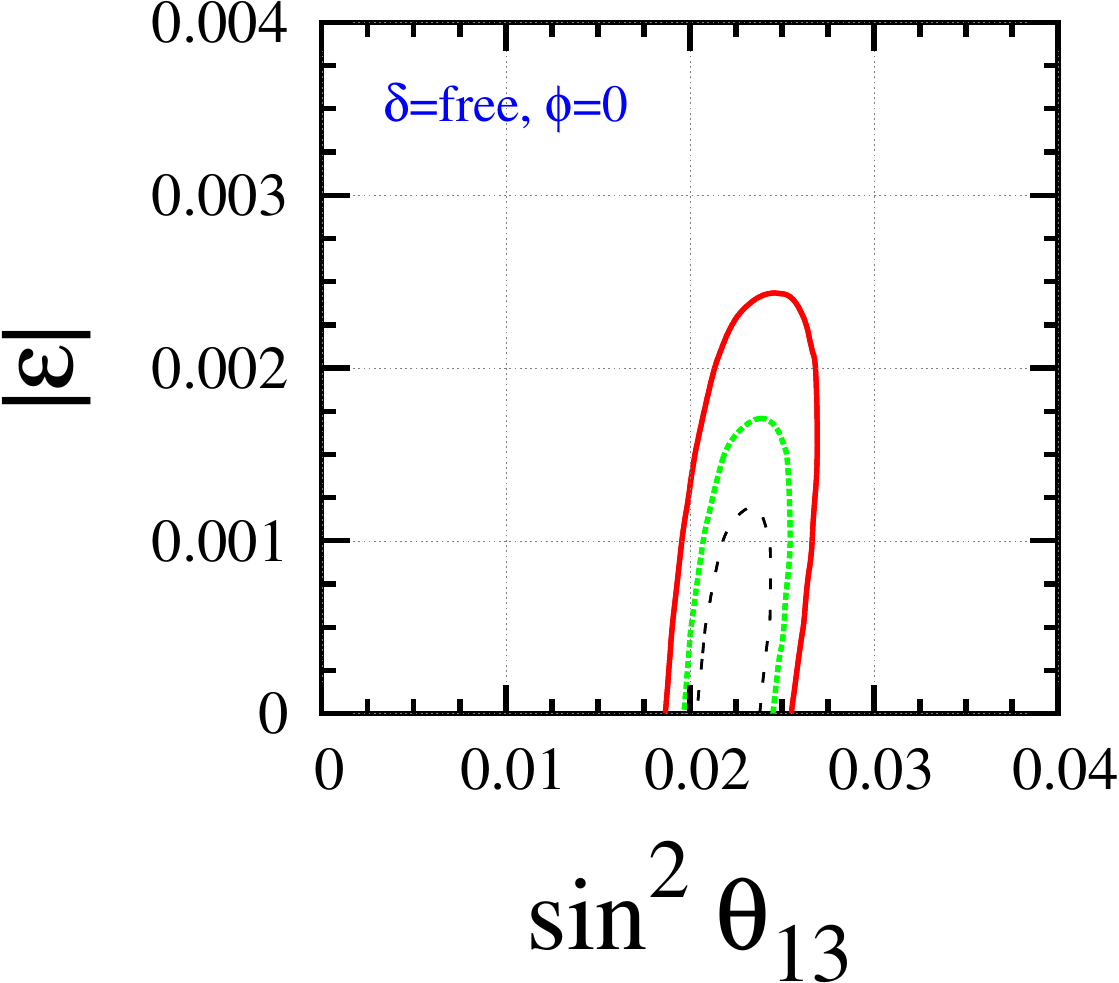}
\includegraphics[width=0.33\textwidth]{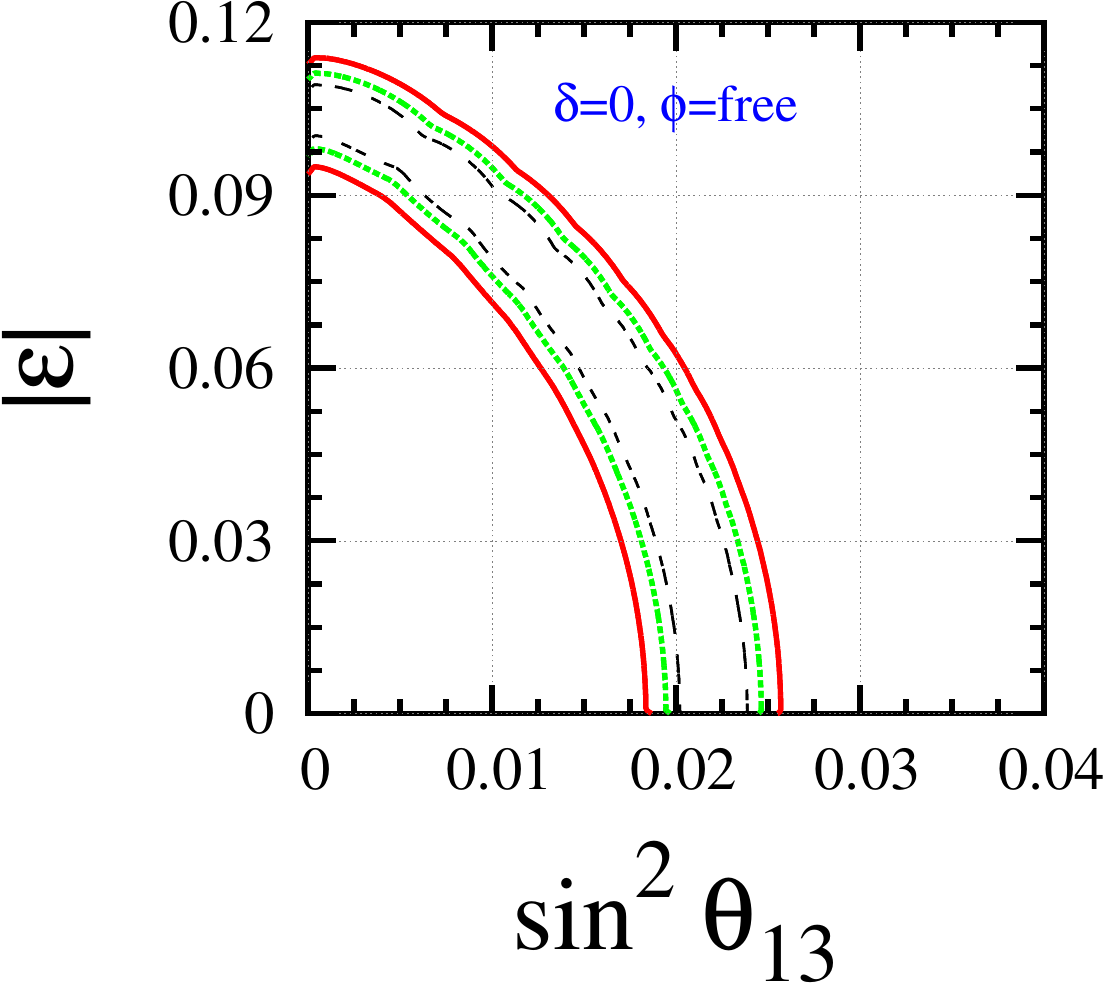}
}

\caption{Allowed region in the $\sin^2\theta_{13}$ - $|\eps|$ plane for different assumptions concerning the phases $\delta$ and $\phi$. The left panel is obtained switching all phases to zero, whereas in the middle panel  $\phi = 0$ and $\delta$ is left free. In the right panel $\delta$ is taken equal to zero while $\phi$ has been marginalized. The conventions for the lines is the same as in Fig.~\ref{fig:eps_e}}
\label{fig:eps_univ}
\end{figure}

Here we present the results obtained under the hypothesis of
flavor-universal NSI, that is, we assume all NSI couplings are present
and they take the same value. Therefore, we consider only two NSI
parameters: $|\eps|$ and $\phi$, in addition to the standard model
parameters entering in the calculations. In this case, the effective
survival probability is given by the expression in
Eq.~(\ref{eq:flavorless}), with separate dependence on the phases
$\delta$ and $\phi$. Therefore we have considered four different cases
in our analysis: one with all the phases set to zero, two cases
varying only one of the phases with the other set to zero and a last
case varying the two phases simultaneously. Our results are presented
at Fig.~\ref{fig:eps_univ}.

The left panel in Fig.~\ref{fig:eps_univ} shows the tight constraint
obtained for the magnitude of the flavor-universal NSI coupling
$|\eps|$ when the phases are set to zero:
\begin{equation}
|\eps| \le 1.2\times 10^{-3}  \quad \, (90\% \, \text{C.L.})\,.
\end{equation}
This result follows directly from the tendency already observed for
$|\eps_e|$ at Fig.~\ref{fig:eps_e}. This happens because, even though
the NSI couplings $|\eps_{\mu,\tau}|$ are also present in the
flavor-universal case, the dominant contribution comes from the
non-oscillatory term depending on $|\eps_e|$.
The same behavior is also present in the middle panel of
Fig.~\ref{fig:eps_univ}, where the Dirac phase $\delta$ is allowed to
freely vary. In this case, the effect of the Dirac phase is just a
scaling in the last term in Eq.~(\ref{eq:shift-univ}), driven by
$\cos{\delta}$, but again the total effect on $\theta_{13}$ is
dominated by the non-oscillatory term in
Eq.~(\ref{eq:shift-univ}). Under this assumption, we obtain the
following bound on the magnitude of the flavor-universal coupling:
\begin{equation}
|\eps| \le 1.3\times 10^{-3}  \quad \, (90\% \, \text{C.L.})\,.
\end{equation}
Note that in the two former cases the allowed range for the mixing
angle $\theta_{13}$ is very close to the one obtained in the standard
case, and therefore the Daya Bay determination is barely affected by
the presence of NSI.
However, when $\delta=0$ and the NSI phase $\phi$ is allowed to take
different values, a completely different behavior results, as it is
shown in the right panel of Fig.~\ref{fig:eps_univ}. In this case, in
the fit of Daya Bay data, the phase $\phi$ takes a value such that the
non-oscillatory term in Eq.~(\ref{eq:flavorless}) is cancelled up to
order $|\eps|^4$.  Most importantly, with a preferred value of
$\cos\phi \simeq -1.5|\eps|$, new terms of second order in $|\eps|$
appear at Eq.~(\ref{eq:flavorless}) and therefore the first order
expression in Eq.~(\ref{eq:shift-univ}) can not satisfactory explain
the degeneracy between $\theta_{13}$ and $|\eps|$. Actually, the shift
in the effective reactor angle is given now by this other expression
(up to order $|\eps|^2$):
\begin{equation}
\tilde{s}^2_{13} 
\approx s^2_{13}+|\varepsilon|^2\left(2 - 3\sqrt{2} s_{13}\right)\,.
\label{eq:shift-univ-fit}
\end{equation} 
Then, it is possible to see that, even with $s_{13} = 0$, one can
reproduce the measured value of $\theta_{13}$ in Daya Bay with $|\eps|
\sim 0.11$, as shown in the corresponding plot.
As a consequence of the degeneracy in the plane $s^2_{13}$ - $|\eps|$,
this specific case is not very restrictive and we get a loose bound on
$|\eps|$:
\begin{equation}
|\eps| \le 1.1\times 10^{-1}  \quad \, (90\% \, \text{C.L.})\,.
\end{equation}
and only an upper bound on $\theta_{13}$:
\begin{equation}
 \sin^2\theta_{13}\le 0.024  \quad \, (90\%  \, \text{C.L.})\,.
\end{equation}
As commented above, the presence of flavor universal NSI implies that
the reactor mixing angle may be compatible with zero. Nevertheless,
the degeneracy between the mixing angle and the new physics parameter
$|\eps|$ observed here may be lifted by a combined analysis with
accelerator long-baseline neutrino experiments.
A global analysis of neutrino data assuming the simultaneous presence
of NSI in reactor and accelerator neutrino data would be very useful
for this purpose.  Besides solving the degeneracy, the combined
analysis might provide further constraints on the NSI couplings as
well as improve the agreement between the preferred $\theta_{13}$
value from reactors and long-baseline experiments, as discussed in
Ref.~\cite{Adhikari:2012vc,Girardi:2014kca,DiIura:2014csa}.
However, since the production, detection and propagation of neutrinos
is quite different in both kind of experiments, a global analysis
would require a very detailed study with many new physics parameters
involved, besides the consideration of a specific model for NSI. In
any case, this point is out of the scope of the present analysis and
it will be considered elsewhere.

Finally, let us comment that we have also considered the case of
flavor-universal NSI with all the phases different from zero. However,
we have not presented the results obtained for this setup because, in
this case, the confusion between $\theta_{13}$ and $|\eps|$ is
complete, and therefore no information on any of the parameters can be
extracted from the analysis of Daya Bay data.
All the results obtained in this section are summarized in Table
\ref{tab:bounds1}.
Needless to mention, to obtain all the limits on $\sin^2\theta_{13}$
presented along this section as well as in the table, we have
marginalized over the NSI couplings over a wide range.  Similarly, to
place bounds on the NSI parameters, $\sin^2\theta_{13}$ has also been
allowed to float over a wide range.

\begin{table}[!t]\centering
\begin{tabular}{|l|c|c|}\hline
phases & $\sin^2\theta_{13}$ & $|\varepsilon|$ \\ 
\hline
\multicolumn{3}{|c|}{electron-type NSI coupling}\\ \hline
$ \phi_e = 0$  & $0.020\le \sin^2\theta_{13}\le 0.024$ &  $|\eps_e| \le 
0.0012$\\ 
$\phi_e$ free  & $0.020\le \sin^2\theta_{13}\le 0.024$ & 
$|\eps_e|$ unbound \\
\hline
\multicolumn{3}{|c|}{muon or tau-type NSI couplings}\\ \hline
$(\delta - \phi_{\mu,\tau}) = 0 $ & $0.013\le \sin^2\theta_{13}\le 0.024$ & 
$|\eps_{\mu,\tau}| \le 0.051$ \\
$(\delta - \phi_{\mu,\tau})$ free & $0.013\le \sin^2\theta_{13}\le 0.036$ & 
$|\eps_{\mu,\tau}| \le 0.052$ \\
\hline
\multicolumn{3}{|c|}{universal NSI couplings}\\ \hline
$\delta = \phi = 0$  & $0.020\le \sin^2\theta_{13}\le 0.024$ & $|\eps| 
\le 0.0012$ \\ \hline
$\delta$ free, $\phi = 0$ & $0.020\le \sin^2\theta_{13}\le 0.025$ & 
$|\eps| \le 0.0013$ \\ \hline
$\delta = 0$, $\phi$ free  & $\sin^2\theta_{13}\le 0.024$ & $|\eps| \le 
0.110$ \\ \hline
\end{tabular}
\caption{\label{tab:bounds1} 90\% C.L. bounds (1 d.o.f) on $\sin^2\theta_{13}$ and the NSI couplings from current Daya Bay data 
without considering any uncertainty in the normalization of reactor event rates in the statistical analysis ($a_\text{norm}$ = 0).
}
\end{table}

\section{Bounds on NSI from  Daya Bay with 5\% normalization error}
\label{NSI-bounds-freenorm}

In the previous section, we have not considered any normalization
error in the statistical analysis of Daya Bay reactor data. This means
that we have assumed a perfect knowledge of the event normalization at
the experiment, disregarding the presence of uncertainties in the flux
reactor normalization or in the detection cross section, among others. 
This procedure has been followed in most of the previous
phenomenological analyses of Daya Bay data in presence of
NSI, see for instance Ref.~\cite{Leitner:2011aa}.
In the more recent work at Ref. ~\cite{Girardi:2014gna}, the authors have considered
small uncertainties in the reactor flux and in the detector properties, although they
did not take into account an uncertainty in the overall normalization of the 
event rates.
Needless to say that a more detailed analysis of reactor data can not 
ignore the presence of such normalization errors. Therefore, in this 
section we present a $\chi^2$ analysis of Daya Bay data using the 
expression defined at Eq.~(\ref{eq:chi2}), where a free normalization 
factor is considered in order to account for the uncertainties in the total
event number normalization.
This point is very relevant in the study of NSI with reactor
experiments, since the uncertainty in the event normalization presents
a degeneracy with the zero-distance effect due to NSI.  In
consequence, the far over near technique exploited by Daya Bay in
order to reduce the dependence upon total normalization does not work
equally fine in the presence of NSI, where the non-oscillatory
zero-distance effect, simultaneously present at near and far
detectors, does not totally cancel.
Actually, in the standard model case without NSI, the number of events
expected at the near detector is given by:
\begin{equation}
N_\text{ND}^\text{SM} \simeq N (1 + a_\text{norm}) P_{ee}^\text{SM}(L=0) = N (1 + a_\text{norm})  \, ,
\end{equation}
while, in the presence of NSI, the event number at the near detector is calculated as follows:
\begin{eqnarray}
N_\text{ND}^\text{NSI} & \simeq & N (1 + a_\text{norm}) P_{ee}^\text{NSI}(L=0) = N (1 + a_\text{norm}) (1+f(\eps)) \nonumber \\
& \simeq & N ( 1 + f(\eps) + a_\text{norm})\, .
\end{eqnarray}
Here $a_\text{norm}$ controls the normalization of far and near
detector events in the fit and, together with the NSI couplings
$\eps$, it fixes the total zero-distance effect.
As a result, if we set $a_\text{norm}$ to zero, we artificially
increase the power of Daya Bay data to constrain the zero-distance
effect due to NSI, getting non-realistic strong bounds on the NSI
couplings.
On the other hand, we can not leave the factor $a_\text{norm}$ totally
free in our statistical analysis, as it is usually done in the
standard Daya Bay analysis, where the factor is kept small thanks to
the far over near technique. Actually, we have found that leaving the
normalization factor totally free, and due to the degeneracy with the
NSI couplings, it could achieve very large values, of the order of
10-20\%. For this reason it is necessary the use of a prior on this
magnitude.
Recent reevaluations of the reactor antineutrino flux indicate an
uncertainty on the total flux of about
3\%~\cite{Mueller:2011nm,Huber:2011wv}. However, an independent
analysis in Ref.~\cite{Hayes:2013wra} claims that this uncertainty may
have been underestimated due to the treatment of forbidden transitions
in the antineutrino flux evaluation, and proposes a total uncertainty
of 4\%. Since the total normalization errors may also include
uncertainties coming from other sources, we follow the conservative
approach of taking a total uncertainty on the reactor event
normalization of 5\%. This is the value we have assumed for $\sigma_a$
in Eq.~(\ref{eq:chi2}). 

\begin{figure}[!tb]
\centerline{
\includegraphics[width=0.33\textwidth]{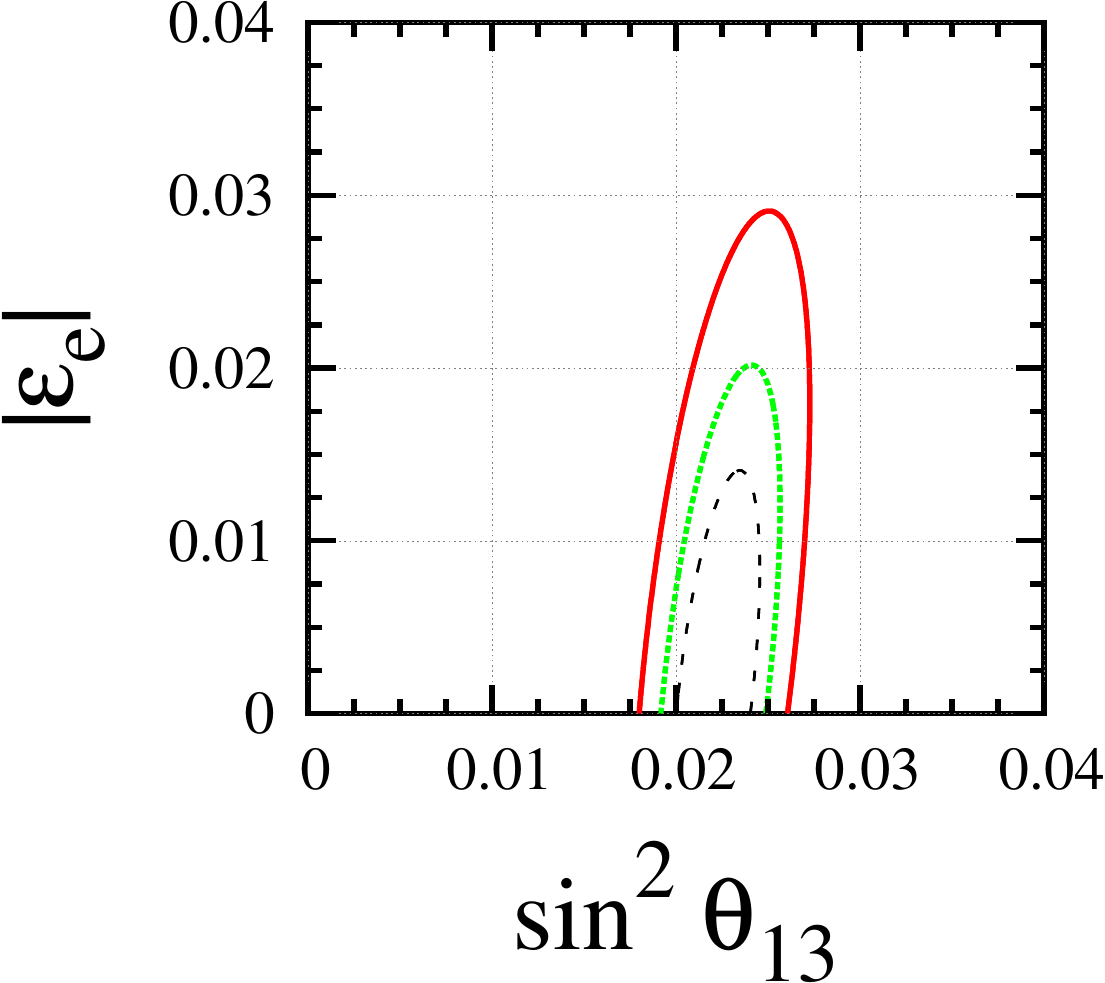}
\includegraphics[width=0.33\textwidth]{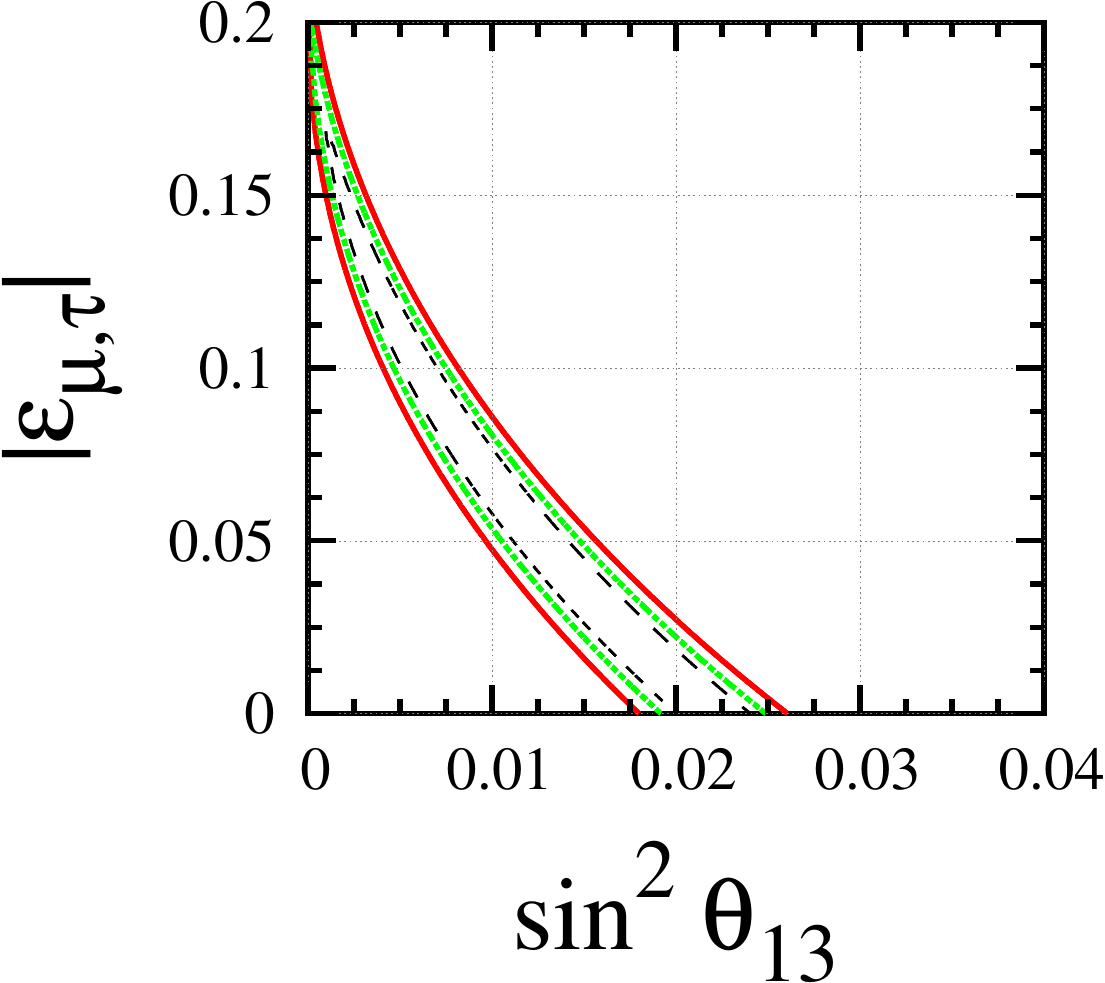}
\includegraphics[width=0.33\textwidth]{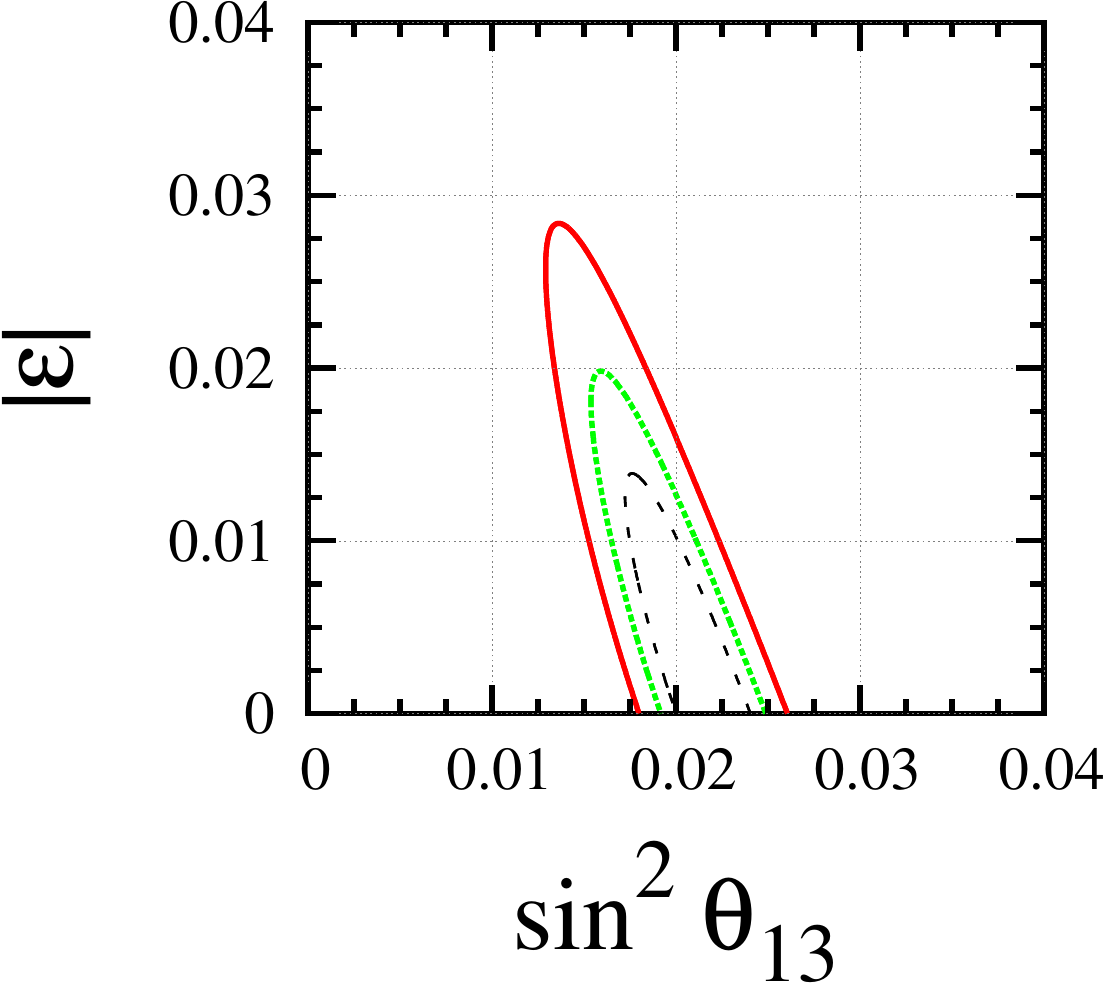}
}
\caption{
Allowed regions in the $\sin^2\theta_{13}$ - NSI coupling plane for the different cases considered: $|\eps_e|$ (left panel), $|\eps_{\mu,\tau}|$ (middle panel) and $|\eps|$ (right panel) setting  all the phases equal to zero and assuming 5\% uncertainty on the total event rate normalization of Daya Bay events. The conventions for the lines is the same as in Fig.~\ref{fig:eps_e}.}
\label{fig:free_norm}
\end{figure}

To illustrate the differences with respect to the results obtained in
the previous section, assuming no uncertainties in the event rate
normalization, here we have considered only the cases where all phases
are set to zero\footnote{Note that, in this case, the correction terms to the effective reactor 
angle at first order in $|\varepsilon|$ dominate and, therefore, second order corrections are not relevant.}. 
The results obtained with these assumptions are
presented in Fig.~\ref{fig:free_norm} and Table \ref{tab:bounds2}.
In the left panel of Fig.~\ref{fig:free_norm} we present the allowed
region in the plane $\sin^2\theta_{13}$ - $|\eps_e|$ when only NSI
with electron antineutrinos are present. In this case, the range for
the reactor mixing angle is rather similar to the one shown in the
left panel of Fig.~\ref{fig:eps_e}:
\begin{equation}
0.020\le \sin^2\theta_{13}\le 0.025 \quad \, (90\% \, \text{C.L.})\, ,
\end{equation}
while the bound on $|\eps_e|$, however, is much weaker than the one
given at Eq.~(\ref{eq:bound_eps_e}) (although still slightly better
than the one at Eq.~(\ref{eq:bound-biggio})):
\begin{equation}
|\eps_e| \le 1.5\times 10^{-2}  \quad \, (90\% \, \text{C.L.})\,.
\label{eq:bound_eps_e-2}
\end{equation}
The same bound is also obtained in the flavor-universal case for the
NSI parameter $|\eps|$, see the right panel of
Fig.~\ref{fig:free_norm}. In this case, the allowed range for the
reactor mixing angle is a bit enlarged with respect to the previous
one:
\begin{equation}
0.017\le \sin^2\theta_{13}\le 0.024 \quad \, (90\% \, \text{C.L.})\, ,
\end{equation}
due to the presence of NSI oscillation terms driven by the new physics
couplings with muon and tau antineutrinos.
As commented above, the loss of sensitivity to the NSI couplings is
due to the degeneracy between the normalization uncertainty and the
zero-distance terms induced by the presence of NSI.  In this way, a
larger value of the NSI parameters can be compensated with a non-zero
normalization factor $a_\text{norm}$, without spoiling the good
agreement with experimental reactor data.

Finally, the middle panel of Fig.~\ref{fig:free_norm} shows the
results obtained when only NSI with muon or tau antineutrinos are
considered. In this case, the cancellation between the normalization
term and the zero-distance effect due to terms of second order in
$|\eps_{\mu,\tau}|$ results in an extended region in the
$\sin^2\theta_{13}$ - $|\eps_{\mu,\tau}|$ plane with an upper bound
of:
\begin{equation}
\sin^2\theta_{13}\le 0.024 \quad \, (90\% \, \text{C.L.})\, .
\end{equation}
The bound on the NSI coupling is given by:
\begin{equation}
|\eps_{\mu,\tau}| \le 0.176 \quad \, (90\% \, \text{C.L.}) \,.
\label{eq:bound_eps_mu-2}
\end{equation}

\begin{table}[!t]\centering
\begin{tabular}{|l|c|c|} \hline
Case & $\sin^2\theta_{13}$ & $|\varepsilon|$ \\ \hline
$\phi_e = 0$ & $0.020\le \sin^2\theta_{13}\le 0.025$ & $|\eps_e|\le 
0.015$ \\ 
$(\delta - \phi_{\mu,\tau}) = 0$ & $\sin^2\theta_{13}\le 0.024$ & 
$|\eps_{\mu,\tau}| \le 0.176$ \\
$\delta = \phi= 0$ & $0.017\le \sin^2\theta_{13}\le 0.024$ & $|\eps| 
\le 0.015$ \\ \hline
\end{tabular}
\caption{\label{tab:bounds2} 90\% C.L. bounds (1 d.o.f) on $\sin^2\theta_{13}$ and the NSI couplings using current Daya Bay data with a 5\% uncertainty on the total event normalization.
}
\end{table}

\section{Comparing NSI constraints from 217 and 621 days of Daya Bay run}
\label{Sec-stat-Daya-Bay}

 \begin{figure}[!tb]
 \centering
 \includegraphics[width=0.49\textwidth]{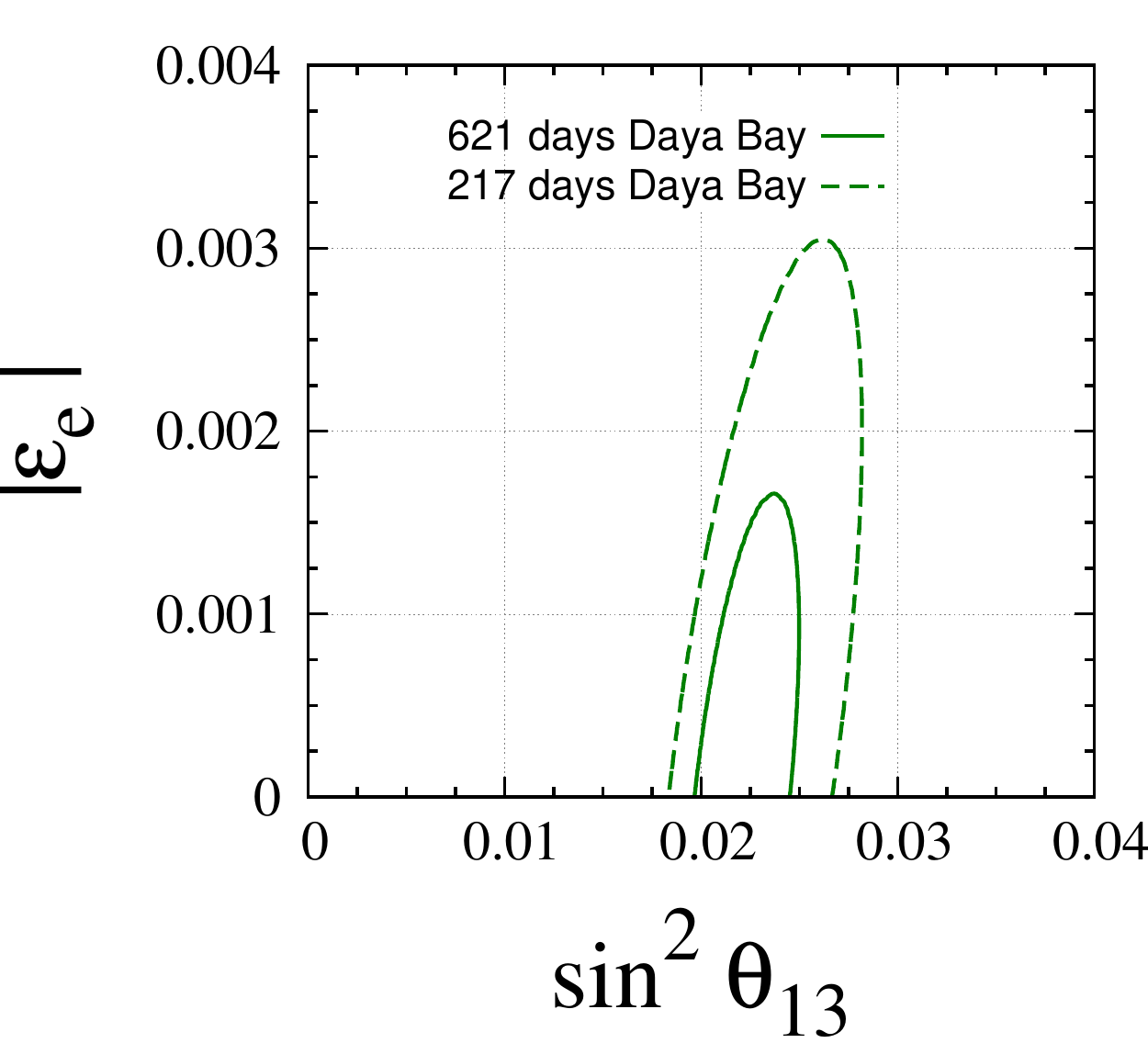}
 \includegraphics[width=0.49\textwidth]{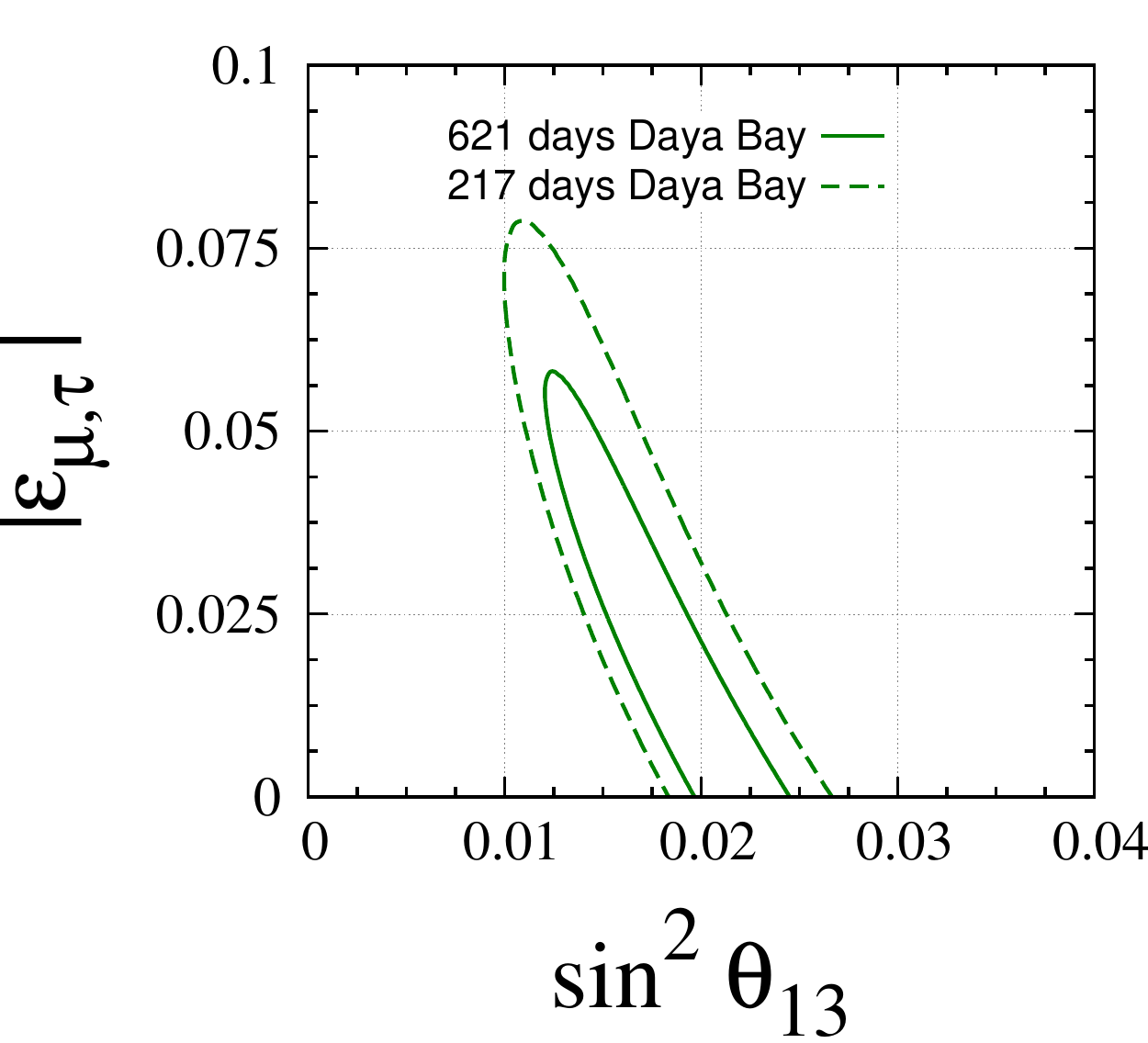}
 \caption{Left panel shows the allowed parameter space in $\sch$ -- $|\eps_e|$ plane at 90\% C.L. (2 d.o.f). 
 Right panel depicts the same in $\sch$ -- $|\eps_{\mu,\tau}|$ plane. Here the solid (dashed) lines correspond 
 to the new (old) 621 (217) days of Daya Bay data. Here all the phases are considered to be zero and 
 the normalization of events is fixed in the statistical analysis with $a_\text{norm}$ = 0.}
 \label{fig:old_new_DB}
 \end{figure}

 The new high-precision data from Daya Bay with 621 days of running
 time \cite{talk-Zhang-Nu2014} has improved the measurement of $\stch$
 dramatically with a relative 1$\sigma$ precision of $\sim$ 6\% as
 compared to $\sim$ 11\% obtained using the previously published 217
 days of data \cite{An:2013zwz}. This clearly shows the impact of the
 four time more statistics that the Daya Bay experiment has
 accumulated with the help of eight ADs in comparison with the
 previously released Daya Bay data set with six ADs. Now it would be
 quite interesting to see how much we can further constrain the
 allowed ranges for these NSI parameters under consideration using the
 new 621 days of Daya Bay data in comparison with the old 217 days of
 Daya Bay run.  In Fig.~\ref{fig:old_new_DB}, we compare the
 performance of the current and the previous data sets of Daya Bay in
 constraining the allowed regions in $\sch$ -- $|\eps_e|$ plane (left
 panel) and in $\sch$ -- $|\eps_{\mu,\tau}|$ plane (right panel). In
 both the panels, the solid (dashed) lines portray the results with
 the new (old) 621 (217) days of Daya Bay data.  For the sake of
 illustration, we have only chosen the cases of NSI parameters which
 are associated with $\anue$ (left panel) and $\anumu$ (right panel)
 assuming all the phases to be zero.  In particular, we have focused
 on the situations which are presented in sections \ref{subsec:nsie}
 and \ref{subsec:nsimu}, where we do not consider the normalization
 uncertainty on the reactor events and set $a_\text{norm}$ = 0 in the
 statistical analysis. As compared to the old data set, with the
 current data, the improvement in constraining the allowed parameter
 space between $\sch$ and the NSI parameters is quite significant as
 can be readily seen from Fig.~\ref{fig:old_new_DB}.

\begin{table}[!t]\centering
\begin{tabular}{|l|c|c|}\hline
Daya Bay data & $\sin^2\theta_{13}$ & $|\varepsilon|$ \\ 
\hline
\multicolumn{3}{|c|}{electron-type NSI parameters} \\ \hline
Current (621 days) & $0.020\le \sin^2\theta_{13}\le 0.024$ &  $|\eps_e| \le 0.0012$ \\ 
Previous (217 days) & $0.019\le \sin^2\theta_{13}\le 0.027$ &  $|\eps_e| \le 0.0024$ \\ 
\hline
\multicolumn{3}{|c|}{muon or tau-type NSI parameters}\\ \hline
Current (621 days) & $0.013\le \sin^2\theta_{13}\le 0.024$ & $|\eps_{\mu,\tau}| \le 0.051$ \\
Previous (217 days) & $0.011\le \sin^2\theta_{13}\le 0.026$ & $|\eps_{\mu,\tau}| \le 0.070$ \\
\hline
\multicolumn{3}{|c|}{universal NSI parameters}\\ \hline
Current (621 days)  & $0.020\le \sin^2\theta_{13}\le 0.024$ & $|\eps| \le 0.0012$ \\ 
Previous (217 days)  & $0.019\le \sin^2\theta_{13}\le 0.026$ & $|\eps| \le 0.0024$ \\
 \hline
\end{tabular}
\caption{\label{tab:bounds3}90\% C.L. (1 d.o.f) bounds on $\sin^2\theta_{13}$ and the NSI parameters 
using the old (new) 217 (621) days of Daya Bay data. Here all the phases are considered to be zero.
We also do not consider the uncertainty on the normalization of reactor events and set $a_\text{norm}$ = 0 
in the statistical analysis.}
\end{table}

In Table~\ref{tab:bounds3}, we present the 90\% C.L. (1 d.o.f)
constraints on $\sin^2\theta_{13}$ and the NSI parameters using the
previous (current) 217 (621) days of Daya Bay data.  Here all the
phases are considered to be zero. We do not consider the normalization
uncertainty on the reactor events and take $a_\text{norm}$ = 0 in the
statistical analysis.
We can see from Table~\ref{tab:bounds3} that in case of electron-type
and universal NSI parameters, the bounds on $|\eps_e|$ and $|\eps|$
are the same while analyzing 621 days of Daya Bay data.  This feature
is also there in the case of 217 days of Daya Bay run.  The
constraints on $|\eps_e|$ and $|\eps|$ get improved by factor of two
when we consider the current 621 days of Daya Bay data compared to its
previous 217 days of data.  We also get better bounds on
$|\eps_{\mu,\tau}|$ using the current Daya Bay data as compared to the
old data set. These results suggest that the future data from the Daya
Bay experiment with more statistics is going to play an important role
to further constrain the allowed parameter space for the NSI
parameters. There are also marginal improvements on the bounds to
$\sin^2\theta_{13}$ with the current 621 days of Daya Bay data.

\begin{table}[!th]\centering
\begin{tabular}{|l|c|c|}\hline
Daya Bay data & $\sin^2\theta_{13}$ & $|\varepsilon|$ \\ 
\hline
\multicolumn{3}{|c|}{electron-type NSI parameters [$\phi_e$ free]} \\ \hline
Current (621 days) & $0.020\le \sin^2\theta_{13}\le 0.024$ & $|\eps_e|$ unbound \\ 
Previous (217 days) & $0.019\le \sin^2\theta_{13}\le 0.027$ & $|\eps_e|$ unbound \\ 
\hline
\multicolumn{3}{|c|}{muon or tau-type NSI parameters [($\delta - \phi_{\mu,\tau}$) free]} \\ \hline
Current (621 days) & $0.013\le \sin^2\theta_{13}\le 0.036$ & $|\eps_{\mu,\tau}| \le 0.052$ \\
Previous (217 days) & $0.011\le \sin^2\theta_{13}\le 0.045$ & $|\eps_{\mu,\tau}| \le 0.070$ \\
\hline
\multicolumn{3}{|c|}{universal NSI parameters [$\delta$ free, $\phi = 0$]} \\ \hline
Current (621 days)  & $0.020\le \sin^2\theta_{13}\le 0.025$ & $|\eps| \le 0.0013$ \\ 
Previous (217 days)  & $0.019\le \sin^2\theta_{13}\le 0.028$ & $|\eps| \le 0.0024$ \\
 \hline
\multicolumn{3}{|c|}{universal NSI parameters [$\delta = 0$, $\phi$ free]} \\ \hline
Current (621 days)  & $\sin^2\theta_{13}\le 0.024$ & $|\eps| \le 0.110$ \\ 
Previous (217 days)  & $\sin^2\theta_{13}\le 0.026$ & $|\eps| \le 0.116$ \\
\hline 
\end{tabular}
\caption{\label{tab:bounds4}90\% C.L. (1 d.o.f) bounds on $\sin^2\theta_{13}$ and the NSI parameters 
using the old (new) 217 (621) days of Daya Bay data. Here we allow the phases or their certain
combinations to vary freely. We do not consider the uncertainty on the normalization of reactor events and 
set $a_\text{norm}$ = 0 in the statistical analysis.}
\end{table}

Finally in Table~\ref{tab:bounds4}, we compare the the 90\% C.L. (1 d.o.f) limits on $\sin^2\theta_{13}$ and 
the NSI parameters obtained using the old 217 days and new 621 days of Daya Bay data allowing 
the phases or their certain combinations to vary freely as we consider in Table~\ref{tab:bounds1} 
in Sec.~\ref{NSI-bounds}. Like in Table~\ref{tab:bounds3}, here also we do not take into account the
normalization uncertainty on the reactor events and consider $a_\text{norm}$ = 0 in the statistical analysis.
Table~\ref{tab:bounds4} depicts that even if we allow the phases to vary freely, we obtain better limits
on the NSI parameters with the new data set as compared to the previous 217 days of Daya Bay 
data, except for the case of $|\eps_e|$ which remains unbounded.
Table~\ref{tab:bounds4} indicates that the new data set of Daya Bay also reduces the allowed ranges for 
$\sin^2\theta_{13}$ in the presence of NSI parameters as compared to the old data. This improvement
is quite significant in the case of $|\eps_{\mu,\tau}|$ even if the effective phase ($\delta - \phi_{\mu,\tau}$)
is allowed to vary freely in the fit.

\section{Summary and conclusions}
\label{Sec-Summary}

The success of the currently running Daya Bay, RENO, and Double Chooz
reactor antineutrino experiments in measuring the smallest lepton
mixing angle $\theta_{13}$ with impressive accuracy signifies an
important advancement in the field of modern neutrino physics with
nonzero mass and three-flavor mixing.  With this remarkable discovery,
the neutrino oscillation physics has entered into a high-precision era
opening up the possibility of observing sub-dominant effects due to
possible new physics beyond the Standard Model of particle physics.
At present, undoubtedly the Daya Bay experiment in China is playing a
leading and an important role in this direction. The recent
high-precision and unprecedentedly copious data from the Daya Bay
experiment has provided us an opportunity to probe the existence of
the non-standard interaction effects which might crop up at the
production point or at the detection stage of the reactor
antineutrinos.

In this paper for the first time, we have reported the new constraints
on the flavor {\it non-universal} and also flavor {\it universal} NSI
parameters obtained using the currently released 621 days of Daya Bay
data. While placing the bounds on these NSI parameters, we have
assumed that the new physics effects are just inverse of each other in
the production and detection processes of the reactor antineutrino
experiment ,{\it i.e.}, $\varepsilon^s_{e\gamma} =
\varepsilon^{d*}_{\gamma e}$.  Considering this special case, we have
discussed in detail the impact of the NSI parameters on the effective
antineutrino survival probability expressions which we ultimately use
to analyze the Daya Bay data. With this special choice of the NSI
parameters, we have observed a shift in the oscillation
amplitude without altering the $L/E$ pattern of the oscillation probability.
This shift in the depth of the oscillation dip can be
caused due to the NSI parameters and as well as $\theta_{13}$, making
it quite difficult to disentangle the NSI effects from the standard
oscillations.  Before presenting the final results, we have studied
the correlations between the NSI parameters and $\theta_{13}$ with the
help of iso-probability surface plots in Sec.~\ref{Reactor-NSI}. This
study has been quite useful to understand the final bounds on the NSI
parameters that we have obtained from the fit.  Since the shape of the
oscillation probability is not distorted with the special choice of
the NSI parameters considered in this paper, an analysis based on the
{\it total event rate} at the Daya Bay experiment is sufficient to
obtain the limits on the NSI parameters.

As far as the flavor {\it non-universal} NSI parameters are concerned,
first we have considered the NSI parameters $|\varepsilon_e|$ and
$\phi_e$ which are associated with $\anue$.  Assuming $\phi_e$ =
0$^\circ$ and a perfect knowledge of the normalization of the event
rates, the current Daya Bay data places a strong constrain on
$|\varepsilon_e| \le 1.2 \times 10^{-3}$ (90\% C.L.) improving the
present bound on $|\varepsilon_e|$ by one order of magnitude. Now if
we consider the uncertainty in the normalization of event rates with a
prior of 5\%, then this limit changes to $15 \times 10^{-3}$, diluting
the constraint by almost one order of magnitude.  No limits can be
placed on $|\varepsilon_e|$ if we allow $\phi_e$ to vary freely in the
fit. We have also observed that the determination of the 1-3 mixing
angle is quite robust in this specific case and it is almost
independent of the issue of uncertainty in the normalization of event
rates and the choice of $\phi_e$. In fact, the allowed range of $\sch$
coincides exactly with the allowed range in absence of NSI. Next we
turn our attention to the NSI parameters $|\varepsilon_\mu|$ and
$\phi_\mu$ which are associated with $\anumu$.  With ($\delta -
\phi_\mu$) = 0$^\circ$ and a perfect knowledge of the event rates
normalization, the current Daya Bay data sets a limit of
$|\varepsilon_\mu| \le 5.1 \times 10^{-2}$ (90\% C.L.)  which is
comparable and complementary to the existing bound obtained using a
different data set under different assumptions. This limit on
$|\varepsilon_\mu|$ becomes $17.6 \times 10^{-2}$ once we consider the
uncertainty in the normalization of event rates with a prior of
5\%. These limits on $|\varepsilon_\mu|$ remain almost unchanged even
if we allow ($\delta - \phi_\mu$) to vary freely in the fit. This is
not true while placing the constraint on $\sch$. For an example, in
the case when we do not consider any uncertainty in the normalization
of event rates, the upper bound on $\sch$ increases from $2.4 \times
10^{-2}$ to $3.6 \times 10^{-2}$ due to freely varying ($\delta -
\phi_\mu$) in the fit instead of setting it to zero.  We cannot place
a lower bound on $\sch$ when we consider the uncertainty in the
normalization of event rates with a prior of 5\% even if ($\delta -
\phi_\mu$) is taken to be zero in the fit.  The above mentioned limits
on $|\varepsilon_\mu|$ and $\sch$ are also valid for the NSI
parameters $|\varepsilon_\tau|$ and $\phi_\tau$ as long as the 2-3
mixing angle is maximal ,{\it i.e.}, $\sa =0.5$.

In the case of flavor {\it universal} NSI parameters, we have placed
limits on $|\eps|$ and $\sch$ under certain assumptions on $\delta$
and $\phi$. With $\delta = \phi = 0^\circ$ and perfect knowledge of
the normalization of the event rates, the upper bound on $|\eps|$ at
90\% C.L. (1 d.o.f) is $1.2 \times 10^{-3}$. Though this limit does
not change much when we marginalize over $\delta$ in the fit, it
deteriorates by almost ninety times when we allow $\phi$ to vary
freely in the fit providing a limit of $|\eps| \le 110 \times 10^{-3}$
(90\% C.L.). With a 5\% uncertainty in the normalization of total
event rates and $\delta = \phi = 0^\circ$, the constraint on $|\eps|$
becomes $\le 15 \times 10^{-3}$ at 90\% C.L. With the same
assumptions, we can restrict $\theta_{13}$ within a range of $0.017
\le \sch \le 0.024$ (90\% C.L.).

One of the novelties of this work is the inclusion of correction terms of 
second order in the NSI couplings $|\varepsilon|$ in the effective 
neutrino probability in Eq.~(\ref{eq:case1}). 
The role of these second order corrections at the effective probability level 
has been analyzed in Sec.~\ref{Special-NSI-Case} as well as in the discussion 
of the probability and correlation plots in Sec.~\ref{probability-vs-energy} and 
Sec.~\ref{iso-probability}.
The impact of second order terms on the determination of $\sch$ is specially 
relevant for the flavour-universal-NSI case, as commented in 
Sec.~\ref{subsec:FU}.

One of the interesting studies that we have performed in this paper is
the comparison of the constraints on the NSI parameters placed with
the current 621 days of Daya Bay data with the limits obtained using
the previously released 217 days of Daya Bay run.  In this analysis,
all the phases are considered to be zero and the normalization of
events is also kept fixed in the statistical analysis with
$a_\text{norm}$ = 0.  We have observed that the constraints on
$|\eps_e|$ and $|\eps|$ get improved by factor of two when we analyze
the current 621 days of Daya Bay data compared to its previous 217
days data.  This comparative study reveals the merit of the huge
statistics that Daya Bay has already accumulated. It also suggests
that the future high-precision data from the Daya Bay experiment with
enhanced statistics is inevitable to further probe the sub-leading
effects in neutrino flavor conversion due to the presence of the
possible NSI parameters beyond the standard three-flavor oscillation
paradigm.

\subsubsection*{Acknowledgments}

S.K.A. acknowledges the support from DST/INSPIRE Research Grant
[IFA-PH-12], Department of Science and Technology, India.
The work of D.V.F. and M.T. was supported by the Spanish grants
FPA2011-22975 and MULTIDARK CSD2009-00064 (MINECO) and
PROMETEOII/2014/084 (Generalitat Valenciana).
This work has also been supported by the U.S.  Department of Energy
under award number DE-SC0003915.

\begin{appendix}

\section{Effective Survival Probability with the NSI parameters:  $\varepsilon^s_{e\gamma} \ne \varepsilon^{d*}_{\gamma e}$}
\label{appendix1}

In this appendix, we present the effective probability expressions for
the physical scenarios where $\varepsilon^s_{e\gamma} \ne
\varepsilon^{d*}_{\gamma e}$. In such cases, the spectral study of the
reactor data plays an important role since the NSI parameters are not
only responsible for a shift in $\theta_{13}$ {\ie} the change of the
{\it depth} of the first oscillation maximum but they also modify the
$L/E$ pattern of the probability due to the shift in its energy.  We have
already mentioned that a detailed analysis of the Daya Bay data
considering such interesting physical cases will be performed in
\cite{shape-nsi-reactor}.

\subsection{Presence of the NSI parameters only at the production stage}

Here we assume that the NSI parameters only affect the production
mechanism of the antineutrinos in the reactor experiment.  It allows
us to write (dropping the universal e-index):
\begin{eqnarray}
 \varepsilon^s_{\gamma} = |\varepsilon_{\gamma}| {\rm e}^{{\rm i} \phi_{\gamma}}  \,, & \text{and} &
 \varepsilon^{d}_{\gamma} = 0 \,.
\end{eqnarray}
With this assumption, we get the effective neutrino survival probability as follows: 
\begin{eqnarray}
P_{\bar{\nu}^s_e \rightarrow \bar{\nu}^d_e} &\simeq& P_{\bar{\nu}^s_e \rightarrow \bar{\nu}^d_e}^\text{SM} +
  |\varepsilon_e|^2 + 2 |\varepsilon_e| \cos \phi_e   \nonumber \\ 
  & + & 2  s_{13} \left[ s_{23} |\varepsilon_\mu|  \sin(\delta-\phi_\mu) + c_{23} |\varepsilon_\tau|  \sin(\delta-\phi_\tau) \right] \,\sin \left(2\Delta_{31}\right) \nonumber \\
  & - & 4 s_{13} \left[ s_{23} |\varepsilon_\mu|  \cos(\delta-\phi_\mu) + c_{23}|\varepsilon_\tau| \cos(\delta-\phi_\tau) \right] \,\sin^2 \left(\Delta_{31}\right)\nonumber \\
 & + & \sin 2\theta_{12} \left[- c_{23} |\varepsilon_\mu| \sin \phi_\mu + s_{23} |\varepsilon_\tau| \sin\phi_\tau \right] \,\sin \left(2\Delta_{21}\right) \,.
\end{eqnarray}

\subsection{NSI at the source and detector with the same magnitude and different phases}

In this case, we assume that the magnitude of the NSI parameters is
the same at the production and detection level, but the phases
associated with the NSI parameters are different at the source and
detector. Under this situation, we can write:
\begin{eqnarray}
 \varepsilon^s_{\gamma} = |\varepsilon_{\gamma}| {\rm e}^{{\rm i} \phi^s_{\gamma}} \,, & \text{and}  &
\varepsilon^{d}_{\gamma} = |\varepsilon_{\gamma}| {\rm e}^{{\rm i} \phi^d_{\gamma}} \,.
\end{eqnarray}
Under this assumption, the effective neutrino survival probability takes the form: 
\begin{eqnarray}
P_{\bar{\nu}_e^s \to \bar{\nu}_e^d} & \simeq &  P^\text{SM}_{\bar{\nu}_e \to
  \bar{\nu}_e}+ P_{\text{non-osc}}^{\text{NSI-IIb}} +  P_{\text{osc-atm}}^{\text{NSI-IIb}}  + P_{\text{osc-solar}}^{\text{NSI-IIb}} \,,
\label{eq:pee_IIb}
\end{eqnarray}
where the non-standard terms are given by:
\begin{eqnarray}
 P_{\text{non-osc}}^{\text{NSI-IIb}}  & = & 2 \left \{ |\eps_e| \left(\cos \phi_e^d +\cos \phi_e^s \right) +|\eps_e|^2 \left[ 1 + \cos (\phi_e^d-\phi_e^s)+\cos(\phi_e^d+\phi_e^s)\right] \right. \nonumber \\
& + & \left. |\eps_\mu|^2 \cos(\phi_\mu^s+\phi_\mu^d)+|\eps_\tau|^2 \cos(\phi_\tau^s+\phi_\tau^d) \right \}    \, ,\\ 
 P_{\text{osc-atm}}^{\text{NSI-IIb}} & = & 2 \left \{ s_{13} s_{23} |\eps_\mu| \left[ \sin(\delta-\phi_\mu^s) - \sin(\delta+\phi_\mu^d)\right]  + s_{13} c_{23} |\eps_\tau| \left[ \sin(\delta-\phi_\tau^s)-\sin(\delta+\phi_\tau^d)\right]   \right. \nonumber \\ 
 & - & s_{23}^2 |\eps_\mu|^2 \sin(\phi_\mu^s+\phi_\mu^d) -  c_{23}^2 |\eps_\tau|^2  \sin(\phi_\tau^s+\phi_\tau^d)  \nonumber \\ & - & \left. c_{23} s_{23} |\eps_\mu| |\eps_\tau| \left[\sin(\phi_\tau^s+\phi_\mu^d) + \sin(\phi_\mu^s+\phi_\tau^d)\right] \right \} \,\sin \left(2\Delta_{31}\right) \nonumber \\
 & - & 4 \left \{   s_{13} s_{23} |\eps_\mu| \left[ \cos(\delta-\phi_\mu^s)+\cos(\delta+\phi_\mu^d)\right] + 
   c_{23} s_{13} |\eps_\tau| \left[ \cos(\delta-\phi_\tau^s)+\cos(\delta+\phi_\tau^d)\right] \right. \nonumber \\
  & + & s_{23}^2 |\eps_\mu|^2 \cos(\phi_\mu^s+\phi_\mu^d) + c_{23}^2|\eps_\tau|^2 \cos(\phi_\tau^s+\phi_\tau^d)  \nonumber \\
              & + & \left. c_{23} s_{23} |\eps_\mu| |\eps_\tau| \left[ \cos(\phi_\tau^s+\phi_\mu^d) + \cos(\phi_\mu^s+\phi_\tau^d) \right] \right \}\,\sin^2 \left(\Delta_{31}\right) \, ,\\ 
 P_{\text{osc-solar}}^{\text{NSI-IIb}}& = & \sin 2\theta_{12} \left[ - c_{23} |\eps_\mu| (\sin \phi_\mu^s+\sin \phi_\mu^d)+ s_{23}|\eps_\tau| (\sin \phi_\tau^s+\sin \phi_\tau^d) \right]\,\sin \left(2\Delta_{21}\right) \,.
\end{eqnarray}

\end{appendix}

\bibliographystyle{JHEP}
\bibliography{nsi-references}

\providecommand{\href}[2]{#2}\begingroup\raggedright\begin{thebibliography}{10}

\bibitem{An:2012bu}
{\bf Daya Bay} Collaboration, F.~An et~al., {\it {Improved Measurement of
  Electron Antineutrino Disappearance at Daya Bay}},  {\em Chin.Phys.} {\bf
  C37} (2013) 011001, [\href{http://xxx.lanl.gov/abs/1210.6327}{{\tt
  arXiv:1210.6327}}].

\bibitem{An:2013zwz}
{\bf Daya Bay} Collaboration, F.~An et~al., {\it {Spectral measurement of
  electron antineutrino oscillation amplitude and frequency at Daya Bay}},
  {\em Phys.Rev.Lett.} {\bf 112} (2014) 061801,
  [\href{http://xxx.lanl.gov/abs/1310.6732}{{\tt arXiv:1310.6732}}].

\bibitem{talk-Zhang-Nu2014}
C.~Zhang, {\it {Recent results from Daya Bay}},  2014.
\newblock Talk given at the Neutrino 2014 Conference, Boston.

\bibitem{Ahn:2012nd}
{\bf RENO} Collaboration, J.~Ahn et~al., {\it {Observation of Reactor Electron
  Antineutrino Disappearance in the RENO Experiment}},  {\em Phys.Rev.Lett.}
  {\bf 108} (2012) 191802, [\href{http://xxx.lanl.gov/abs/1204.0626}{{\tt
  arXiv:1204.0626}}].

\bibitem{Forero:2014bxa}
D.~V. Forero, M.~T\'ortola, and J.~W.~F. Valle, {\it Neutrino oscillations
  refitted},  {\em Phys. Rev.} {\bf D90} (2014) 093006,
  [\href{http://xxx.lanl.gov/abs/1405.7540}{{\tt arXiv:1405.7540}}].

\bibitem{Capozzi:2013csa}
F.~Capozzi, G.~Fogli, E.~Lisi, A.~Marrone, D.~Montanino, et~al., {\it {Status
  of three-neutrino oscillation parameters, circa 2013}},  {\em Phys.Rev.} {\bf
  D89} (2014) 093018, [\href{http://xxx.lanl.gov/abs/1312.2878}{{\tt
  arXiv:1312.2878}}].

\bibitem{Gonzalez-Garcia:2014bfa}
M.~Gonzalez-Garcia, M.~Maltoni, and T.~Schwetz, {\it {Updated fit to three
  neutrino mixing: status of leptonic CP violation}},
  \href{http://xxx.lanl.gov/abs/1409.5439}{{\tt arXiv:1409.5439}}.

\bibitem{Mohapatra:2006gs}
R.~Mohapatra and A.~Smirnov, {\it {Neutrino Mass and New Physics}},  {\em
  Ann.Rev.Nucl.Part.Sci.} {\bf 56} (2006) 569--628,
  [\href{http://xxx.lanl.gov/abs/hep-ph/0603118}{{\tt hep-ph/0603118}}].

\bibitem{Morisi:2012fg}
S.~Morisi and J.~Valle, {\it {Neutrino masses and mixing: a flavour symmetry
  roadmap}},  {\em Fortsch.Phys.} {\bf 61} (2013) 466--492,
  [\href{http://xxx.lanl.gov/abs/1206.6678}{{\tt arXiv:1206.6678}}].

\bibitem{King:2014nza}
S.~F. King, A.~Merle, S.~Morisi, Y.~Shimizu, and M.~Tanimoto, {\it {Neutrino
  Mass and Mixing: from Theory to Experiment}},  {\em New J.Phys.} {\bf 16}
  (2014) 045018, [\href{http://xxx.lanl.gov/abs/1402.4271}{{\tt
  arXiv:1402.4271}}].

\bibitem{Abe:2011fz}
{\bf Double Chooz} Collaboration, Y.~Abe et~al., {\it {Indication for the
  disappearance of reactor electron antineutrinos in the Double Chooz
  experiment}},  {\em Phys.Rev.Lett.} {\bf 108} (2012) 131801,
  [\href{http://xxx.lanl.gov/abs/1112.6353}{{\tt arXiv:1112.6353}}].

\bibitem{Abe:2012tg}
{\bf Double Chooz} Collaboration, Y.~Abe et~al., {\it {Reactor electron
  antineutrino disappearance in the Double Chooz experiment}},  {\em Phys.Rev.}
  {\bf D86} (2012) 052008, [\href{http://xxx.lanl.gov/abs/1207.6632}{{\tt
  arXiv:1207.6632}}].

\bibitem{Adamson:2013ue}
{\bf MINOS} Collaboration, P.~Adamson et~al., {\it {Electron neutrino and
  antineutrino appearance in the full MINOS data sample}},  {\em
  Phys.Rev.Lett.} (2013) [\href{http://xxx.lanl.gov/abs/1301.4581}{{\tt
  arXiv:1301.4581}}].

\bibitem{Abe:2013hdq}
{\bf T2K} Collaboration, K.~Abe et~al., {\it {Observation of Electron Neutrino
  Appearance in a Muon Neutrino Beam}},  {\em Phys.Rev.Lett.} {\bf 112} (2014)
  061802, [\href{http://xxx.lanl.gov/abs/1311.4750}{{\tt arXiv:1311.4750}}].

\bibitem{Abe:2013xua}
{\bf T2K} Collaboration, K.~Abe et~al., {\it {Evidence of Electron Neutrino
  Appearance in a Muon Neutrino Beam}},  {\em Phys.Rev.} {\bf D88} (2013)
  032002, [\href{http://xxx.lanl.gov/abs/1304.0841}{{\tt arXiv:1304.0841}}].

\bibitem{Nunokawa:2007qh}
H.~Nunokawa, S.~J. Parke, and J.~W. Valle, {\it {CP Violation and Neutrino
  Oscillations}},  {\em Prog.Part.Nucl.Phys.} {\bf 60} (2008) 338--402,
  [\href{http://xxx.lanl.gov/abs/0710.0554}{{\tt arXiv:0710.0554}}].

\bibitem{Pascoli:2013wca}
S.~Pascoli and T.~Schwetz, {\it {Prospects for neutrino oscillation physics}},
  {\em Adv.High Energy Phys.} {\bf 2013} (2013) 503401.

\bibitem{Agarwalla:2013hma}
S.~K. Agarwalla, S.~Prakash, and S.~U. Sankar, {\it {Exploring the three flavor
  effects with future superbeams using liquid argon detectors}},
  \href{http://xxx.lanl.gov/abs/1304.3251}{{\tt arXiv:1304.3251}}.

\bibitem{Agarwalla:2014fva}
S.~K. Agarwalla, {\it {Physics Potential of Long-Baseline Experiments}},  {\em
  Adv.High Energy Phys.} {\bf 2014} (2014) 457803,
  [\href{http://xxx.lanl.gov/abs/1401.4705}{{\tt arXiv:1401.4705}}].

\bibitem{Minakata:2014tza}
H.~Minakata, {\it {Neutrino Physics Now and in the Near Future}},
  \href{http://xxx.lanl.gov/abs/1403.3276}{{\tt arXiv:1403.3276}}.

\bibitem{Minkowski:1977sc}
P.~Minkowski, {\it {mu $\rightarrow$ e gamma at a Rate of One Out of 1-Billion
  Muon Decays?}},  {\em Phys.Lett.} {\bf B67} (1977) 421.

\bibitem{Yanagida:1979as}
T.~Yanagida, {\it {Horizontal symmetry and masses of neutrinos}},  {\em
  Conf.Proc.} {\bf C7902131} (1979) 95--99.

\bibitem{Mohapatra:1979ia}
R.~N. Mohapatra and G.~Senjanovic, {\it {Neutrino Mass and Spontaneous Parity
  Violation}},  {\em Phys.Rev.Lett.} {\bf 44} (1980) 912.

\bibitem{GellMann:1980vs}
M.~Gell-Mann, P.~Ramond, and R.~Slansky, {\it {Complex Spinors and Unified
  Theories}},  {\em Conf.Proc.} {\bf C790927} (1979) 315--321,
  [\href{http://xxx.lanl.gov/abs/1306.4669}{{\tt arXiv:1306.4669}}].

\bibitem{Schechter:1980gr}
J.~Schechter and J.~Valle, {\it {Neutrino Masses in SU(2) x U(1) Theories}},
  {\em Phys.Rev.} {\bf D22} (1980) 2227.

\bibitem{Lazarides:1980nt}
G.~Lazarides, Q.~Shafi, and C.~Wetterich, {\it {Proton Lifetime and Fermion
  Masses in an SO(10) Model}},  {\em Nucl.Phys.} {\bf B181} (1981) 287--300.

\bibitem{Mohapatra:1986bd}
R.~Mohapatra and J.~Valle, {\it {Neutrino Mass and Baryon Number
  Nonconservation in Superstring Models}},  {\em Phys.Rev.} {\bf D34} (1986)
  1642.

\bibitem{Akhmedov:1995ip}
E.~K. Akhmedov, M.~Lindner, E.~Schnapka, and J.~Valle, {\it {Left-right
  symmetry breaking in NJL approach}},  {\em Phys.Lett.} {\bf B368} (1996)
  270--280, [\href{http://xxx.lanl.gov/abs/hep-ph/9507275}{{\tt
  hep-ph/9507275}}].

\bibitem{Akhmedov:1995vm}
E.~K. Akhmedov, M.~Lindner, E.~Schnapka, and J.~Valle, {\it {Dynamical
  left-right symmetry breaking}},  {\em Phys.Rev.} {\bf D53} (1996) 2752--2780,
  [\href{http://xxx.lanl.gov/abs/hep-ph/9509255}{{\tt hep-ph/9509255}}].

\bibitem{Dev:2009aw}
P.~B. Dev and R.~Mohapatra, {\it {TeV Scale Inverse Seesaw in SO(10) and
  Leptonic Non-Unitarity Effects}},  {\em Phys.Rev.} {\bf D81} (2010) 013001,
  [\href{http://xxx.lanl.gov/abs/0910.3924}{{\tt arXiv:0910.3924}}].

\bibitem{Boucenna:2014zba}
S.~M. Boucenna, S.~Morisi, and J.~W. Valle, {\it {The low-scale approach to
  neutrino masses}},  {\em Adv.High Energy Phys.} {\bf 2014} (2014) 831598,
  [\href{http://xxx.lanl.gov/abs/1404.3751}{{\tt arXiv:1404.3751}}].

\bibitem{Cheng:1980qt}
T.~Cheng and L.-F. Li, {\it {Neutrino Masses, Mixings and Oscillations in SU(2)
  x U(1) Models of Electroweak Interactions}},  {\em Phys.Rev.} {\bf D22}
  (1980) 2860.

\bibitem{Zee:1980ai}
A.~Zee, {\it {A Theory of Lepton Number Violation, Neutrino Majorana Mass, and
  Oscillation}},  {\em Phys.Lett.} {\bf B93} (1980) 389.

\bibitem{Babu:1988ki}
K.~Babu, {\it {Model of 'Calculable' Majorana Neutrino Masses}},  {\em
  Phys.Lett.} {\bf B203} (1988) 132.

\bibitem{Diaz:1997xc}
M.~A. Diaz, J.~C. Romao, and J.~Valle, {\it {Minimal supergravity with R-parity
  breaking}},  {\em Nucl.Phys.} {\bf B524} (1998) 23--40,
  [\href{http://xxx.lanl.gov/abs/hep-ph/9706315}{{\tt hep-ph/9706315}}].

\bibitem{Hirsch:2000ef}
M.~Hirsch, M.~Diaz, W.~Porod, J.~Romao, and J.~Valle, {\it {Neutrino masses and
  mixings from supersymmetry with bilinear R parity violation: A Theory for
  solar and atmospheric neutrino oscillations}},  {\em Phys.Rev.} {\bf D62}
  (2000) 113008, [\href{http://xxx.lanl.gov/abs/hep-ph/0004115}{{\tt
  hep-ph/0004115}}].

\bibitem{Roulet:1991sm}
E.~Roulet, {\it {MSW effect with flavor changing neutrino interactions}},  {\em
  Phys.Rev.} {\bf D44} (1991) 935--938.

\bibitem{Guzzo:1991hi}
M.~Guzzo, A.~Masiero, and S.~Petcov, {\it {On the MSW effect with massless
  neutrinos and no mixing in the vacuum}},  {\em Phys.Lett.} {\bf B260} (1991)
  154--160.

\bibitem{Barger:1991ae}
V.~D. Barger, R.~Phillips, and K.~Whisnant, {\it {Solar neutrino solutions with
  matter enhanced flavor changing neutral current scattering}},  {\em
  Phys.Rev.} {\bf D44} (1991) 1629--1643.

\bibitem{Bergmann:1999pk}
S.~Bergmann, Y.~Grossman, and D.~M. Pierce, {\it {Can lepton flavor violating
  interactions explain the atmospheric neutrino problem?}},  {\em Phys.Rev.}
  {\bf D61} (2000) 053005, [\href{http://xxx.lanl.gov/abs/hep-ph/9909390}{{\tt
  hep-ph/9909390}}].

\bibitem{Berezhiani:2001rs}
Z.~Berezhiani and A.~Rossi, {\it {Limits on the nonstandard interactions of
  neutrinos from e+ e- colliders}},  {\em Phys.Lett.} {\bf B535} (2002)
  207--218, [\href{http://xxx.lanl.gov/abs/hep-ph/0111137}{{\tt
  hep-ph/0111137}}].

\bibitem{Antusch:2008tz}
S.~Antusch, J.~P. Baumann, and E.~Fernandez-Martinez, {\it {Non-Standard
  Neutrino Interactions with Matter from Physics Beyond the Standard Model}},
  {\em Nucl.Phys.} {\bf B810} (2009) 369--388,
  [\href{http://xxx.lanl.gov/abs/0807.1003}{{\tt arXiv:0807.1003}}].

\bibitem{Gavela:2008ra}
M.~Gavela, D.~Hernandez, T.~Ota, and W.~Winter, {\it {Large gauge invariant
  non-standard neutrino interactions}},  {\em Phys.Rev.} {\bf D79} (2009)
  013007, [\href{http://xxx.lanl.gov/abs/0809.3451}{{\tt arXiv:0809.3451}}].

\bibitem{Malinsky:2008qn}
M.~Malinsky, T.~Ohlsson, and H.~Zhang, {\it {Non-Standard Neutrino Interactions
  from a Triplet Seesaw Model}},  {\em Phys.Rev.} {\bf D79} (2009) 011301,
  [\href{http://xxx.lanl.gov/abs/0811.3346}{{\tt arXiv:0811.3346}}].

\bibitem{Ohlsson:2009vk}
T.~Ohlsson, T.~Schwetz, and H.~Zhang, {\it {Non-standard neutrino interactions
  in the Zee-Babu model}},  {\em Phys.Lett.} {\bf B681} (2009) 269--275,
  [\href{http://xxx.lanl.gov/abs/0909.0455}{{\tt arXiv:0909.0455}}].

\bibitem{Kopp:2007ne}
J.~Kopp, M.~Lindner, T.~Ota, and J.~Sato, {\it {Non-standard neutrino
  interactions in reactor and superbeam experiments}},  {\em Phys.Rev.} {\bf
  D77} (2008) 013007, [\href{http://xxx.lanl.gov/abs/0708.0152}{{\tt
  arXiv:0708.0152}}].

\bibitem{Leitner:2011aa}
R.~Leitner, M.~Malinsky, B.~Roskovec, and H.~Zhang, {\it {Non-standard
  antineutrino interactions at Daya Bay}},  {\em JHEP} {\bf 1112} (2011) 001,
  [\href{http://xxx.lanl.gov/abs/1105.5580}{{\tt arXiv:1105.5580}}].

\bibitem{Girardi:2014gna}
I.~Girardi and D.~Meloni, {\it {Constraining new physics scenarios in neutrino
  oscillations from Daya Bay data}},
  \href{http://xxx.lanl.gov/abs/1403.5507}{{\tt arXiv:1403.5507}}.

\bibitem{Biggio:2009nt}
C.~Biggio, M.~Blennow, and E.~Fernandez-Martinez, {\it {General bounds on
  non-standard neutrino interactions}},  {\em JHEP} {\bf 0908} (2009) 090,
  [\href{http://xxx.lanl.gov/abs/0907.0097}{{\tt arXiv:0907.0097}}].

\bibitem{Berezhiani:2001rt}
Z.~Berezhiani, R.~Raghavan, and A.~Rossi, {\it {Probing nonstandard couplings
  of neutrinos at the Borexino detector}},  {\em Nucl.Phys.} {\bf B638} (2002)
  62--80, [\href{http://xxx.lanl.gov/abs/hep-ph/0111138}{{\tt
  hep-ph/0111138}}].

\bibitem{Miranda:2004nb}
O.~Miranda, M.~Tortola, and J.~Valle, {\it {Are solar neutrino oscillations
  robust?}},  {\em JHEP} {\bf 0610} (2006) 008,
  [\href{http://xxx.lanl.gov/abs/hep-ph/0406280}{{\tt hep-ph/0406280}}].

\bibitem{Barranco:2005ps}
J.~Barranco, O.~Miranda, C.~Moura, and J.~Valle, {\it {Constraining
  non-standard interactions in nu(e) e or anti-nu(e) e scattering}},  {\em
  Phys.Rev.} {\bf D73} (2006) 113001,
  [\href{http://xxx.lanl.gov/abs/hep-ph/0512195}{{\tt hep-ph/0512195}}].

\bibitem{Barranco:2007ej}
J.~Barranco, O.~Miranda, C.~Moura, and J.~Valle, {\it {Constraining
  non-standard neutrino-electron interactions}},  {\em Phys.Rev.} {\bf D77}
  (2008) 093014, [\href{http://xxx.lanl.gov/abs/0711.0698}{{\tt
  arXiv:0711.0698}}].

\bibitem{Bolanos:2008km}
A.~Bolanos, O.~Miranda, A.~Palazzo, M.~Tortola, and J.~Valle, {\it {Probing
  non-standard neutrino-electron interactions with solar and reactor
  neutrinos}},  {\em Phys.Rev.} {\bf D79} (2009) 113012,
  [\href{http://xxx.lanl.gov/abs/0812.4417}{{\tt arXiv:0812.4417}}].

\bibitem{Escrihuela:2009up}
F.~Escrihuela, O.~Miranda, M.~Tortola, and J.~Valle, {\it {Constraining
  nonstandard neutrino-quark interactions with solar, reactor and accelerator
  data}},  {\em Phys.Rev.} {\bf D80} (2009) 105009,
  [\href{http://xxx.lanl.gov/abs/0907.2630}{{\tt arXiv:0907.2630}}].

\bibitem{Davidson:2003ha}
S.~Davidson, C.~Pena-Garay, N.~Rius, and A.~Santamaria, {\it {Present and
  future bounds on nonstandard neutrino interactions}},  {\em JHEP} {\bf 0303}
  (2003) 011, [\href{http://xxx.lanl.gov/abs/hep-ph/0302093}{{\tt
  hep-ph/0302093}}].

\bibitem{Escrihuela:2011cf}
F.~Escrihuela, M.~Tortola, J.~Valle, and O.~Miranda, {\it {Global constraints
  on muon-neutrino non-standard interactions}},  {\em Phys.Rev.} {\bf D83}
  (2011) 093002, [\href{http://xxx.lanl.gov/abs/1103.1366}{{\tt
  arXiv:1103.1366}}].

\bibitem{Forero:2011zz}
D.~Forero and M.~Guzzo, {\it {Constraining nonstandard neutrino interactions
  with electrons}},  {\em Phys.Rev.} {\bf D84} (2011) 013002.

\bibitem{Fornengo:2001pm}
N.~Fornengo, M.~Maltoni, R.~Tomas, and J.~Valle, {\it {Probing neutrino
  nonstandard interactions with atmospheric neutrino data}},  {\em Phys.Rev.}
  {\bf D65} (2002) 013010, [\href{http://xxx.lanl.gov/abs/hep-ph/0108043}{{\tt
  hep-ph/0108043}}].

\bibitem{Gonzalez-Garcia:2013usa}
M.~Gonzalez-Garcia and M.~Maltoni, {\it {Determination of matter potential from
  global analysis of neutrino oscillation data}},  {\em JHEP} {\bf 1309} (2013)
  152, [\href{http://xxx.lanl.gov/abs/1307.3092}{{\tt arXiv:1307.3092}}].

\bibitem{Choubey:2014iia}
S.~Choubey and T.~Ohlsson, {\it {Bounds on Non-Standard Neutrino Interactions
  Using PINGU}},  \href{http://xxx.lanl.gov/abs/1410.0410}{{\tt
  arXiv:1410.0410}}.

\bibitem{Li:2014mlo}
Y.-F. Li and Y.-L. Zhou, {\it {Shifts of neutrino oscillation parameters in
  reactor antineutrino experiments with non-standard interactions}},  {\em
  Nucl.Phys.} {\bf B888} (2014) 137--153,
  [\href{http://xxx.lanl.gov/abs/1408.6301}{{\tt arXiv:1408.6301}}].

\bibitem{Ohlsson:2008gx}
T.~Ohlsson and H.~Zhang, {\it {Non-Standard Interaction Effects at Reactor
  Neutrino Experiments}},  {\em Phys.Lett.} {\bf B671} (2009) 99--104,
  [\href{http://xxx.lanl.gov/abs/0809.4835}{{\tt arXiv:0809.4835}}].

\bibitem{Bilenky:1992wv}
S.~M. Bilenky and C.~Giunti, {\it {Seesaw type mixing and muon-neutrino to
  tau-neutrino oscillations}},  {\em Phys.Lett.} {\bf B300} (1993) 137--140,
  [\href{http://xxx.lanl.gov/abs/hep-ph/9211269}{{\tt hep-ph/9211269}}].

\bibitem{Grossman:1995wx}
Y.~Grossman, {\it {Nonstandard neutrino interactions and neutrino oscillation
  experiments}},  {\em Phys.Lett.} {\bf B359} (1995) 141--147,
  [\href{http://xxx.lanl.gov/abs/hep-ph/9507344}{{\tt hep-ph/9507344}}].

\bibitem{GonzalezGarcia:2001mp}
M.~Gonzalez-Garcia, Y.~Grossman, A.~Gusso, and Y.~Nir, {\it {New CP violation
  in neutrino oscillations}},  {\em Phys.Rev.} {\bf D64} (2001) 096006,
  [\href{http://xxx.lanl.gov/abs/hep-ph/0105159}{{\tt hep-ph/0105159}}].

\bibitem{Meloni:2009cg}
D.~Meloni, T.~Ohlsson, W.~Winter, and H.~Zhang, {\it {Non-standard interactions
  versus non-unitary lepton flavor mixing at a neutrino factory}},  {\em JHEP}
  {\bf 1004} (2010) 041, [\href{http://xxx.lanl.gov/abs/0912.2735}{{\tt
  arXiv:0912.2735}}].

\bibitem{Antusch:2006vwa}
S.~Antusch, C.~Biggio, E.~Fernandez-Martinez, M.~Gavela, and J.~Lopez-Pavon,
  {\it {Unitarity of the Leptonic Mixing Matrix}},  {\em JHEP} {\bf 0610}
  (2006) 084, [\href{http://xxx.lanl.gov/abs/hep-ph/0607020}{{\tt
  hep-ph/0607020}}].

\bibitem{shape-nsi-reactor}
S.~K. Agarwalla, D.~V. Forero, and M.~T\'{o}rtola, ``{Spectral Analysis of Daya
  Bay data to investigate non-standard interactions}.'' Work in progress, 2015.

\bibitem{Langacker:1988up}
P.~Langacker and D.~London, {\it {Lepton Number Violation and Massless
  Nonorthogonal Neutrinos}},  {\em Phys.Rev.} {\bf D38} (1988) 907.

\bibitem{Tortola:2012te}
D.~Forero, M.~Tortola, and J.~Valle, {\it {Global status of neutrino
  oscillation parameters after Neutrino-2012}},  {\em Phys.Rev.} {\bf D86}
  (2012) 073012, [\href{http://xxx.lanl.gov/abs/1205.4018}{{\tt
  arXiv:1205.4018}}].

\bibitem{FernandezMartinez:2007ms}
E.~Fernandez-Martinez, M.~Gavela, J.~Lopez-Pavon, and O.~Yasuda, {\it
  {CP-violation from non-unitary leptonic mixing}},  {\em Phys.Lett.} {\bf
  B649} (2007) 427--435, [\href{http://xxx.lanl.gov/abs/hep-ph/0703098}{{\tt
  hep-ph/0703098}}].

\bibitem{Goswami:2008mi}
S.~Goswami and T.~Ota, {\it {Testing non-unitarity of neutrino mixing matrices
  at neutrino factories}},  {\em Phys.Rev.} {\bf D78} (2008) 033012,
  [\href{http://xxx.lanl.gov/abs/0802.1434}{{\tt arXiv:0802.1434}}].

\bibitem{Luo:2008vp}
S.~Luo, {\it {Non-unitary deviation from the tri-bimaximal lepton mixing and
  its implications on neutrino oscillations}},  {\em Phys.Rev.} {\bf D78}
  (2008) 016006, [\href{http://xxx.lanl.gov/abs/0804.4897}{{\tt
  arXiv:0804.4897}}].

\bibitem{Forero:2011pc}
D.~Forero, S.~Morisi, M.~Tortola, and J.~Valle, {\it {Lepton flavor violation
  and non-unitary lepton mixing in low-scale type-I seesaw}},  {\em JHEP} {\bf
  1109} (2011) 142, [\href{http://xxx.lanl.gov/abs/1107.6009}{{\tt
  arXiv:1107.6009}}].

\bibitem{Kopeikin:2004cn}
V.~Kopeikin, L.~Mikaelyan, and V.~Sinev, {\it {Reactor as a source of
  antineutrinos: Thermal fission energy}},  {\em Phys.Atom.Nucl.} {\bf 67}
  (2004) 1892--1899, [\href{http://xxx.lanl.gov/abs/hep-ph/0410100}{{\tt
  hep-ph/0410100}}].

\bibitem{Mueller:2011nm}
T.~Mueller, D.~Lhuillier, M.~Fallot, A.~Letourneau, S.~Cormon, et~al., {\it
  {Improved Predictions of Reactor Antineutrino Spectra}},  {\em Phys.Rev.}
  {\bf C83} (2011) 054615, [\href{http://xxx.lanl.gov/abs/1101.2663}{{\tt
  arXiv:1101.2663}}].

\bibitem{Abazajian:2012ys}
K.~Abazajian, M.~Acero, S.~Agarwalla, A.~Aguilar-Arevalo, C.~Albright, et~al.,
  {\it {Light Sterile Neutrinos: A White Paper}},
  \href{http://xxx.lanl.gov/abs/1204.5379}{{\tt arXiv:1204.5379}}.

\bibitem{Vogel:1999zy}
P.~Vogel and J.~F. Beacom, {\it {Angular distribution of neutron inverse beta
  decay, $\bar\nu_e + p \to e^+ + n$}},  {\em Phys.Rev.} {\bf D60} (1999)
  053003, [\href{http://xxx.lanl.gov/abs/hep-ph/9903554}{{\tt
  hep-ph/9903554}}].

\bibitem{Adhikari:2012vc}
R.~Adhikari, S.~Chakraborty, A.~Dasgupta, and S.~Roy, {\it {Non-standard
  interaction in neutrino oscillations and recent Daya Bay, T2K experiments}},
  {\em Phys.Rev.} {\bf D86} (2012) 073010,
  [\href{http://xxx.lanl.gov/abs/1201.3047}{{\tt arXiv:1201.3047}}].

\bibitem{Girardi:2014kca}
I.~Girardi, D.~Meloni, and S.~Petcov, {\it {The Daya Bay and T2K results on
  $\sin^2 2 \theta_{13}$ and Non-Standard Neutrino Interactions}},  {\em
  Nucl.Phys.} {\bf B886} (2014) 31--42,
  [\href{http://xxx.lanl.gov/abs/1405.0416}{{\tt arXiv:1405.0416}}].

\bibitem{DiIura:2014csa}
A.~Di~Iura, I.~Girardi, and D.~Meloni, {\it {Probing new physics scenarios in
  accelerator and reactor neutrino experiments}},
  \href{http://xxx.lanl.gov/abs/1411.5330}{{\tt arXiv:1411.5330}}.

\bibitem{Huber:2011wv}
P.~Huber, {\it {On the determination of anti-neutrino spectra from nuclear
  reactors}},  {\em Phys.Rev.} {\bf C84} (2011) 024617,
  [\href{http://xxx.lanl.gov/abs/1106.0687}{{\tt arXiv:1106.0687}}].

\bibitem{Hayes:2013wra}
A.~Hayes, J.~Friar, G.~Garvey, G.~Jungman, and G.~Jonkmans, {\it {Systematic
  Uncertainties in the Analysis of the Reactor Neutrino Anomaly}},  {\em
  Phys.Rev.Lett.} {\bf 112} (2014) 202501,
  [\href{http://xxx.lanl.gov/abs/1309.4146}{{\tt arXiv:1309.4146}}].

\end{thebibliography}\endgroup

\end{document}